\documentclass[letterpaper]{article} 
\usepackage{aaai2026}
\usepackage{times}  
\usepackage{helvet}  
\usepackage{courier}  
\usepackage[hyphens]{url}  
\usepackage{graphicx} 
\usepackage{natbib}  
\usepackage{caption} 
\usepackage{algorithm}
\usepackage{algorithmic}
\usepackage{booktabs}
\usepackage{multirow}
\usepackage{amsfonts}
\usepackage{amsmath}

\usepackage{times}
\usepackage{helvet}
\usepackage{courier}
\usepackage{xcolor}

\usepackage{newfloat}
\usepackage{listings}
\DeclareCaptionStyle{ruled}{labelfont=normalfont,labelsep=colon,strut=off} 
\floatstyle{ruled}
\newfloat{listing}{tb}{lst}{}
\floatname{listing}{Listing}
\title{CoMET: A Contrastive-Masked Brain Foundation Model for Universal EEG Representation}
\author{
    Ang Li\textsuperscript{\rm 1, 2, 3}\equalcontrib,
    Zikai Wang\textsuperscript{\rm 1, 3}\equalcontrib,
    Liuyin Yang\textsuperscript{\rm 2},
    Zhenyu Wang\textsuperscript{\rm 1},
    Tianheng Xu\textsuperscript{\rm 1, 3},\\
    Honglin Hu\textsuperscript{\rm 1, 3},
    Marc M. Van Hulle\textsuperscript{\rm 2}
}
\affiliations{
    \textsuperscript{\rm 1}Shanghai Advanced Research Institute, Chinese Academy of Sciences\\

    \textsuperscript{\rm 2}Faculty of Medicine, KU Leuven,\\
    \textsuperscript{\rm 3}University of Chinese Academy of Sciences\\
    (huhl@sari.ac.cn)
}
\nocopyright

\usepackage{bibentry}

\begin{document}

\maketitle

\begin{abstract}
Electroencephalography (EEG) is a non-invasive technique for recording brain activity, widely used in brain-computer interfaces, clinic, and healthcare. Traditional EEG deep models typically focus on specific dataset and task, limiting model size and generalization. Recently, self-supervised brain foundation models have emerged and been applied to various downstream tasks. Nevertheless, these models still have limitations: current SOTA models typically rely on masked reconstruction strategy; however, EEG features of adjacent channels are highly correlated, which causes the pre-training to overly focus on low-dimensional signal-similarity features in local regions and neglect the global discriminative patterns vital for downstream tasks. To address these limitations, we propose a brain foundation model called CoMET. Specifically, we employ the masked autoencoder with redesigned patching and embedding for EEG as backbone and devise a novel contrastive learning framework with mirror-scale augmentation to strengthen the global discrimination ability. CoMET is pre-trained on mixed EEG datasets over 3000 subjects with over one million samples. It is evaluated on ten different downstream datasets, and the SOTA results demonstrate CoMET's superior ability in extracting universal EEG representations and strong clinical potential.
\end{abstract}

\begin{figure}[!t]
\centering
\includegraphics[width=0.88\columnwidth]{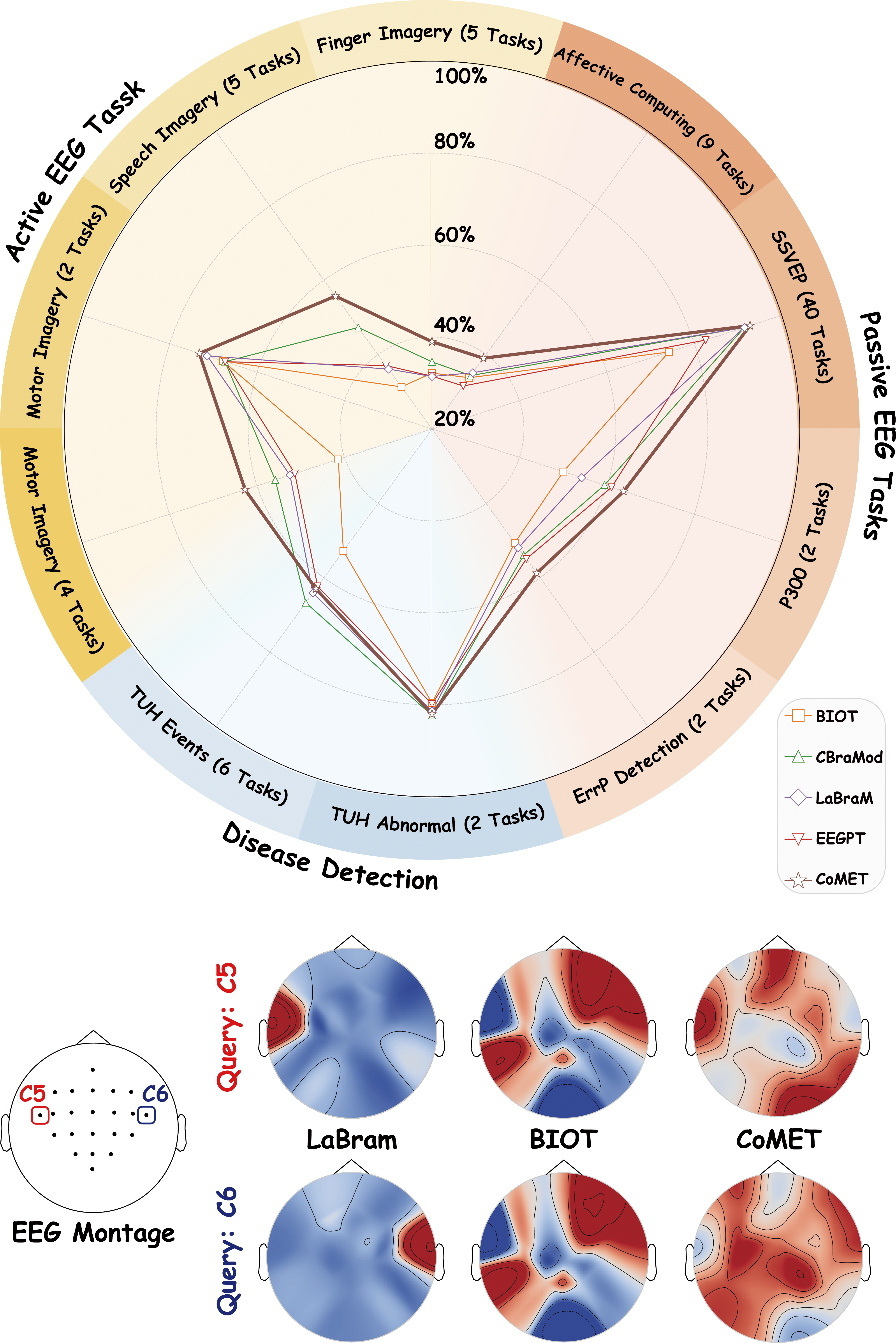}
\caption{\textbf{Top}: Balanced accuracy on ten downstream datasets. \textbf{Bottom}: Self-attention maps on channel-level scalp plots. LabraM (MEM, ICLR'2024) mainly focuses on local area with similar signal of query channel. BIOT (CL, NeurIPS'2023) captures global brain information but collapses into homogeneous attention for different query channels. CoMET (ours) focus on whole-brain in a balanced way with attention diversity.}
\label{fig1}
\vspace{-6pt}
\end{figure}

\section{Introduction}
Electroencephalography (EEG) is a non-invasive technique that captures the amplitude of electrical signals through electrodes placed on the scalp~\cite{teplan2002fundamentals}.
Due to its non-invasive nature and portability, EEG plays a vital role in brain-computer interfaces (BCIs) and healthcare~\cite{edelman2019noninvasive}. EEG signals can be formulated as a matrix $S \in \mathbb{R}^{C \times T}$~\cite{jianglarge}, where $C$ represents the number of channels that could vary across devices, and $T$ represents the sample length that depends on the sampling rate and the collection time. By analyzing the features of EEG, the BCI system can obtain a person's intention, reaction and mental state, such as motor imagery~\cite{al2021deep}, visual evoked potential classification~\cite{norcia2015steady}, emotion recognition~\cite{li2022eeg}, and seizure detection~\cite{chen2025long}, etc.

The EEG signal originates from the cerebral cortex, but the signal is demoted by the intracranial fluid, skull and scalp, causing the signal-to-noise ratio to be low~\cite{nunez2006electric}. To decode EEG, researchers have designed various machine learning and deep learning models, such as FBCSP~\cite{ang2008filter}, EEGNet~\cite{lawhern2018eegnet}, EEGConformer~\cite{song2022eeg}, etc. These models typically focus on EEG samples from specific tasks and devices. When the task and/or device changes, the model needs to be retrained. Moreover, the size of a single dataset is limited when originating from a targeted EEG experiment. It necessitates constrained parameters to avoid overfitting, hindering its capacity to learn robust and generalizable EEG representations~\cite{wang2024eegpt}.

Recently, inspired by transformer-based~\cite{vaswani2017attention} self-supervised learning (SSL) on natural language processing (NLP)~\cite{devlin2019bert} and computer vision (CV)~\cite{baobeit, he2022masked}, some studies have proposed brain foundation models that are pre-trained on a vast amount of heterogeneous EEG data and further adapted to downstream datasets of different tasks. These methods typically slice EEG sample into patches according to the channel-time dimension, add positional embeddings and feed them into a transformer to learn generalizable representations of EEG signals. The SSL strategies of these models fall into two categories. The first comprises contrastive learning (CL), such as BENDR~\cite{kostas2021bendr} and BIOT~\cite{yang2023biot}, which learn view-invariant EEG representations by minimizing the distance between positive pairs and maximizing the separation from negative pairs in the representation space. The second is masked EEG modeling (MEM), such as LabraM~\cite{jianglarge}, EEGPT~\cite{wang2024eegpt}, CBraMod~\cite{wang2025cbramod}, which randomly masks a portion of EEG patches and learn the shallow features by reconstructing the original signals or features through visible patches. Although both strategies successfully distill generic EEG representations across diverse datasets and tasks, several issues remain as follows.

\textbf{Problems:} As discussed in the existing study~\cite{parkself}, CL mainly operates at ``sample-level'' and captures global patterns, but the homogeneous token representations make the encoder suffer from attention collapse, resulting the requirement of high-quality data to converge. On the other hand, MEM functions at the ``token level'' and emphasizes local shallow features. Although this makes it easier to converge during pre-training, it struggles to model global discriminative representations. These issues are even more pronounced when working with EEG data. The EEG signal is temporally non-stationary, making it easier for CL to overfit to task-irrelevant features due to limited-data-induced attention collapse. In contrast, because of volume conduction in EEG, signals recorded at neighbouring electrodes are sourced from similar neurons. Electrodes spaced less than 10 cm apart exhibit high zero-lag correlations across virtually all frequency bands~\cite{brunner2016volume}, which leads the MEM-based methods always focus on neighbouring visible channels' tokens, learn low-dimensional local similarity features to complete the pre-training tasks, and reduce the ability to represent the global discrimination that is important for downstream tasks. \textbf{Figure \ref{fig1}} provides an intuitive visualization of the attention differences between the two strategies (more visualization comparison please refer to Appendix 8).

Based on the above analysis, a plausible idea is to leverage CL to enhance MEM's ability to capture global patterns. Some studies of vision models~\cite{huang2023contrastive} have explored constructing positive and negative pairs through pixel shifting, and the integration of CL and mask image modeling shows better performance than using either alone. However, EEG data differ from images in many aspects, making it more challenging to construct contrastive pairs:
\begin{itemize}
\item EEG has fixed channel locations. Spatial shifting will disrupt the spatial channel dependence. 
\item In contrast to images, cross-sample channel similarity in EEG is consistent. Temporal shifting or shuffle operations rarely change the similarity.
\item The channel dependency of EEG signals leads to performance degradation when a single linear transformation is applied across different channel combinations for building contrastive pairs.
\end{itemize}

To address these issues, we propose a new EEG pre-training framework that integrates strengths of each strategy. Our approach integrates the following components: mirror-scale augmentation, contrastive global representation, channel-time decoupling embedding, and masked token reconstruction inspired by MAE~\cite{he2022masked}.

Based on this framework, we train a large brain foundation model, called CoMET (Contrastive Masked Encoding Transformer) for universal EEG feature extraction. CoMET is pre-trained on over one million EEG samples from different BCI datasets. For the downstream task, we adopt the linear probing strategy that freezes the encoder and evaluate the CoMET on ten popular BCI and medical datasets. Our experiment demonstrates CoMET's state-of-the-art performance in extracting local and global discriminative EEG features for downstream tasks. The contributions of this paper are as follows:
\begin{itemize}
    \item Providing a 151-million-parameter brain foundation model for various BCIs and clinical applications.
    \item Recognizing the challenge of EEG neighboring channels similarity for MEM, and designing mirror-scale augmentation and contrastive global representation, effectively to strengthen the global discrimination ability.
    \item Adoption of MAE-inspired asymmetric encoder and decoder framework that redesigns patching and embedding for EEG, further enhancing the model's ability to capture local features, and demonstration on ten downstream datasets of different tasks and heterogeneous data formats, including different fine-tuning strategies. The results demonstrate that CoMET outperforms other SOTA models, and the performance improvement complies with the scaling law.
\end{itemize}

\begin{figure*}[!h]
\centering
\includegraphics[width=0.92\textwidth]{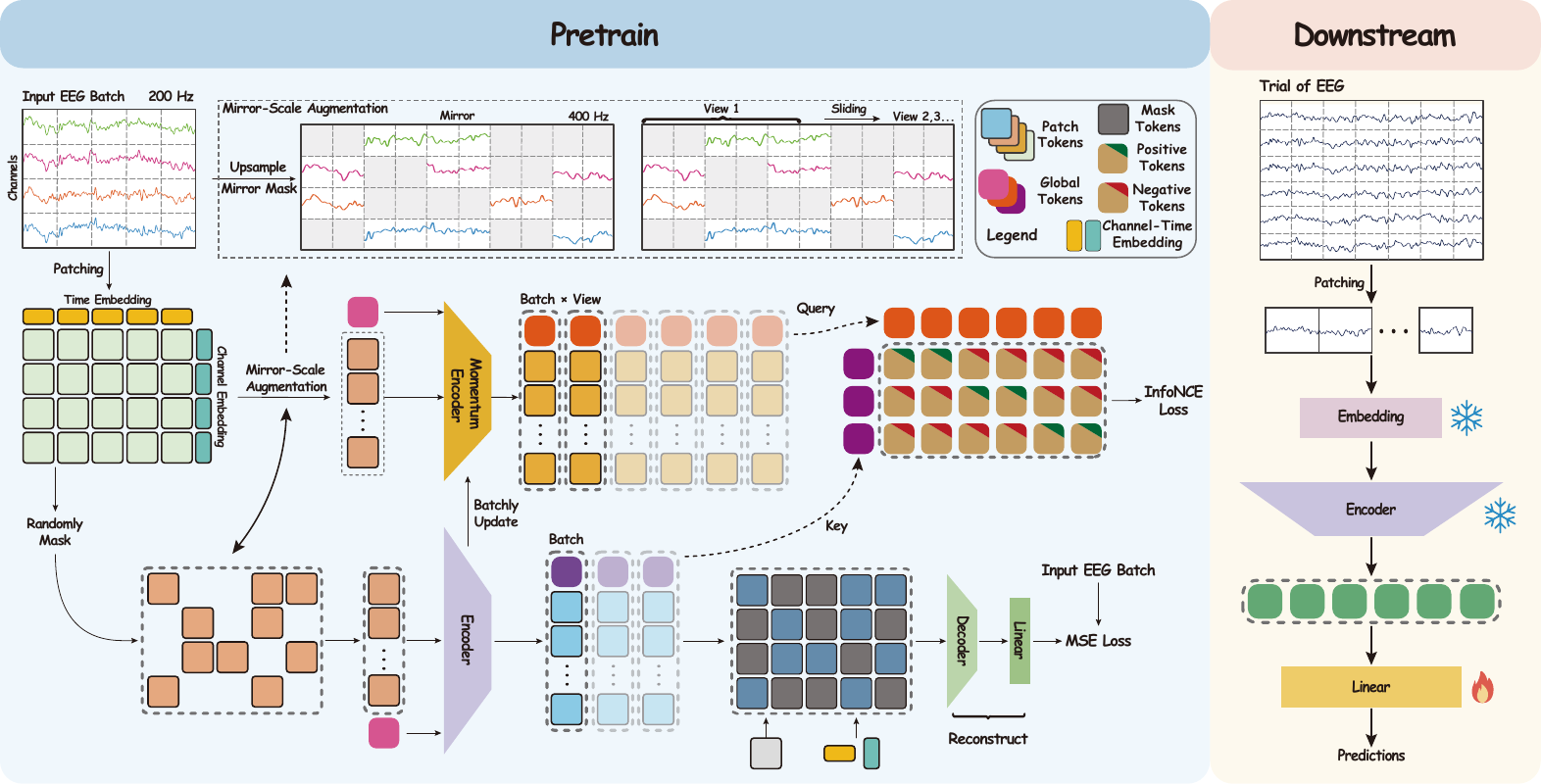}
\caption{The structure of CoMET. (1) \textbf{Pre-training Stage}: There are two branches during the pre-training stage. \textbf{Upper} is the contrastive branch. The momentum encoder receives mirror-scale augmented views (mirror masked compared to input of reconstruction branch) and outputs global tokens. \textbf{Lower} is the reconstruction branch. Raw EEG signals are patched, added channel-temporal embeddings and randomly masked 50\% channels tokens in each time step. Then the visible tokens are sent to encoder together with global token. The output patch tokens are appended with learnable mask tokens and forwarded into decoder to compute the MSE Loss with original signals. The InfoNCE loss is computed between the global tokens of the MEM and momentum encoders, where positive pairs are from the same sample and negative pairs are from different samples in the same batch. (2) \textbf{Downstream Stage}: The pre-trained encoder is frozen and adapted to different tasks via linear probing.}
\label{fig2}
\vspace{-8pt}
\end{figure*}

\section{Related Work}
The brain foundation model is trained on a large amount of EEG data, typically through self-supervised learning, and can be applied on various downstream tasks. BENDR~\cite{kostas2021bendr} trains a Transformer encoder with an InfoNCE contrastive objective on temporally cropped views of the same recording, yielding sample-invariant representations that generalise well across tasks but suffering attention collapse on fine-grained patterns. BIOT~\cite{yang2023biot} extended this paradigm to a multi-biosignal scenario, enriching inputs with explicit channel \& time-position embeddings so the model can align unseen heterogeneous datasets, but it still relies solely on view consistency and faces attention collapse. Afterwards, researchers introduced the masked image modeling~\cite{baobeit,he2022masked} in CV, to make the model converge faster and avoid attention collapse. LabraM~\cite{jianglarge}, inspired by BEiT~\cite{baobeit}, first learns the EEG tokenizer by predicting the frequency domain information and subsequently reconstructs the features of occluded patches via a masked reconstruction strategy. EEGPT~\cite{wang2024eegpt} adopts dual self-supervision by combining spatio-temporal alignment and masked reconstruction, but its pre-trained model only captures spatial features and the temporal dependencies are not utilized for downstream. CBraMod~\cite{wang2025cbramod}, also based on MEM, devises a criss-cross transformer as the backbone and use the asymmetric convolutional positional encoding scheme to encode spatial-temporal positional information.

\section{Methodology}
In this section, we give a detailed description of the proposed CoMET model (\textbf{Figure \ref{fig2}}). The pre-training stage contains two components. \textbf{Masked EEG Modeling}. Departing from previous brain foundation models LabraM, EEGPT and CBraMod, our method employs an asymmetric encoder–decoder design inspired by MAE, which introduces mask tokens only in the decoder phase. The pipeline consists of patching, channel-time embedding, randomly masking, the transformer encoder, double embedding, and the decoder. The goal is to recover the original masked signal. \textbf{Contrastive Learning}.
Generating positive views of input samples through mirror-scale augmentation, and a learnable global token is used to aggregate discriminative representations from visible patches, followed by the InfoNCE loss to optimize contrastive learning between positive and negative features. For the downstream stage, the linear probing is used to effectively evaluate the cross-domain feature extraction capability of the pre-trained model.

\subsection{Patching and Channel-Time Embedding}
Given an EEG sample represented as $S\in \mathbb{R}^{C \times T}$, where $C$ is the number of channels and $T$ the sample length. We first partition $S$ into channel-temporal patches. This is achieved by applying a one-dimensional convolution with same kernel size and a stride of length $l$. It produces non-overlapping vectors as follows:
\begin{equation}
\{x_{i,j}\in \mathbb{R}^{d}|i\in(1, 2,...,C), j\in(1,2,...,N)\},
\end{equation}
where $d$ denotes the embedding dimension and $N=T/l$ the number of temporal patches. For channel $i$ and temporal patch $j$, the token embedding is 
\begin{equation}
e_{i,j} = x_{i,j} + e^c_i + e^t_j,
\end{equation}
where the learnable channel embeddings $\{e_i^{c}\}_{i=1}^{C}\subset\mathbb{R}^d$ are looked up from channel names, and the learnable temporal embeddings $\{e_j^{t}\}_{j=1}^{N}\subset\mathbb{R}^d$ are added in temporal order.
The complete set of patch tokens is
\begin{equation}
    E = \{e_{i,j}|i\in(1, 2,...,C), j\in(1,2,...,N)\}.
\end{equation}

\subsection{Masking}
During the encoding phase, we randomly keep half of the patches at each time position and remove the remaining patches rather than replacing them with masked tokens (mask ratio=0.5, according to CBraMod). The sampling patches keep the original channel and temporal embedding, and each time position has the same number of patches from different channels:
\begin{equation}
    E_v = \{e_{i,j} | i\in C_{j}, j\in(1,2,...,N) \},
\end{equation}
where $C_{j}$ is the visible channel at each time position.

\subsection{MEM Encoder and Decoder}
Before feeding visible patches $E_v$ to the encoder, we append a learnable global token \(e_g\in\mathbb{R}^d\) to summarize global discriminative representations. Unlike vision model, which derives linear projection after the final layer, global token participates in every self-attention layer. The reorganized paches is defined as $\widetilde{E} = E_v \cup {\{e_g\in\mathbb{R}^d\}}$.

Then, visible patches and the global token are fed into the encoder:

\begin{equation}
    F^{(n)} = Attention(\widetilde{E}W_n^Q, \widetilde{E}W_n^K, \widetilde{E}W_n^V),
\end{equation}

where $F_n$ is the $n$-th head attention. The final output of encoder is denoted as:
\begin{equation}
E_{R}
    =\{\,e^{r}_{i,j}
              \;\big|\;
              i \in C_{j},\;
              j = 1,\dots,N
      \} \;\cup\; 
    \{\,e^{r}_{g}\}.
\end{equation}
We use output patch embeddings to reconstruct masked patches, and use the global token embedding $e_g^r$ for discriminative training through CL in the following part. We insert the learnable masked tokens $e_m \in \mathbb{R}^d$ at every masked position. Because the channel-time embeddings of the visible tokens have already been updated by the MEM encoder, we re-add the learnable channel (\(e^{c}_{i}\)) and
temporal (\(e^{t}_{j}\)) positional embeddings to each token. Then all tokens are fed into the transformer decoder:

\begin{equation}
\label{eq:decoder}
\begin{aligned}
\{rec_{i,j}\}
&=\operatorname{Decoder}\Bigl(
      \bigl\{\,e^{r}_{i,j}+e^{c}_{i}+e^{t}_{j}\;\bigl|\;(i,j)\in \mathcal{V}\bigr\} \\
&\quad\cup\,
      \bigl\{\,e_{m}+e^{c}_{i}+e^{t}_{j}\;\bigl|\;(i,j)\in \mathcal{M}\bigr\}
  \Bigr),
\end{aligned}
\end{equation}
where $\mathcal{V}$ and $\mathcal{M}$ denote the index sets of visible and masked tokens,
respectively, and $rec_{c,t}$ the reconstruction of the patch
located at channel \(i\) and time step \(j\).

\subsection{Momentum Encoder}
The momentum encoder~\cite{he2020momentum} is introduced to generate contrastive targets for the MEM encoder \(\mathcal{F}_{o}\) to learn global EEG representations. The momentum encoder shares the same
architecture as \(\mathcal{F}_{o}\) but operates on the mirror-scale augmentation signal.

The inputs of the MEM encoder and momentum encoder can be considered as two perspectives of the brain state. Contrastive learning on these paired views strengthen the encoder’s ability to derive global brain-state features and prepares the encoder for downstream tasks that involve multiple channel combinations.

To focus the contrastive loss on global semantics, we retain only
the global tokens from the momentum encoder and discard its patch
tokens, thereby reducing the similarity among adjacent patches.The momentum encoder $\mathcal{F}_{m}$ is updated through exponential moving average (EMA). Denoting the parameters of $\mathcal{F}_{m}$ and $\mathcal{F}_{o}$ as $\theta_m$ and $\theta_o$, the update rule is 
\begin{equation}
    \theta_m \leftarrow \mu \theta_o + (1-\mu)\theta_m.
\end{equation}

\subsection{Mirror-Scale Augmentation}
Typically, two distinct views of the same input are required in contrastive learning. Because the MEM branch already provides one view, we construct a complementary view to form positive and negative pairs.

We propose a mirror-scale augmentation that (i) preserves global correlations with the original view and avoids token-level similarity that could cause feature collapse and (ii) Generates more paired samples. Let the visible channels for MEM be
\begin{equation}
    V_{a} \;=\;
    \{\,v_{i,j}\ |\; i\!\in\!V^j,\ j=1,\dots,N \}.
\end{equation}
And the visible channels fed to the momentum encoder are defined as:
\begin{equation}
V_{b} = \{v_{i,j}|i\in\{C_{all} \setminus V^j\}, j=1,\dots,N\}. 
\end{equation}
Then the input signal is upsampled from the original sampling rate~\(f\) to
\(2f\), doubling the number of temporal patches from \(N\) to \(2N\) while
keeping the patch length and visible-channel list unchanged. And the visible patch set after upsampling becomes:
\begin{equation}
V_{b}^{(2f)} = \left\{\, v_{i,j} \;\middle|\; i \in \{ C_{\mathrm{all}} \setminus V^{\lceil j/2 \rceil}\},\; j = 1,\dots, 2N \right\}.
\end{equation}
Afterwards, a sliding window of length~\(N\) with stride 1 is applied along the temporal axis, which generates a group of different augmentation views for each sample. The augmented views from the same sample as the MEM encoder are treated as positive pairs, and the augmented view of different samples in the same batch are treated as negative pairs.

\section{Training objective}
\subsection{Pre-training Stage}
Our pre-training objective combines a reconstructive loss that restores masked EEG patches and a contrastive loss that aligns global representations across augmented views.
Given the set of masked indices \(\mathcal{M}\), the reconstructive loss is calculated through mean squared error (MSE):
\begin{equation}
\mathcal{L}_R = \frac{1}{|\mathcal{M}|} \sum_{(i,j) \in \mathcal{M}} \left\| \textit{rec}_{i,j}-raw_{i,j} \right\|_2^2.
\end{equation}

For the contrastive loss, we employ the InfoNCE loss~\cite{oord2018representation}, which simultaneously pulls positive view from the same sample closer while pushing away negative views from different samples. Let $e_g^1$ and $e_g^2$ denote the global tokens produced by the MEM encoder and the momentum encoder, respectively. Their cosine similarity is
\begin{equation}
\rho = \frac{\langle e_g^1, e_g^2\rangle}
            {\lVert e_g^1\rVert_{2}\,\lVert e_g^2\rVert_{2}}.
\end{equation}

For each sample we treat \(\rho^{+}\) (similarity between the two views of the same sample) as the positive pair, and \(\rho^{-}_{i}\) as the negative pair obtained by matching the MEM-view of one sample with the momentum-view of other samples in the same batch. The loss is defined as:
\begin{equation}
\mathcal{L}_C = -\log \frac{\exp(\rho^+ / \tau)}{\exp(\rho^+ / \tau) + \sum_{i=1}^{K-1} \exp(\rho_i^{-} / \tau)}
\end{equation}
where $\tau$ is the temperature. The total loss is calculated by combining $\mathcal{L}_R$ and $\mathcal{L}_C$.

\subsection{Downstream Stage}
For downstream stage, we adopt the linear-probing strategy~\cite{alain2016understanding}: the parameters of the pre-trained encoder are kept frozen, and only a lightweight linear head is updated. All the patch embeddings produced by the encoder are fed to task-specific linear layer. In this way, the downstream performance directly reflects the quality of the representations learned during pre-training. In addition, we also conduct full parameter fine-tuning experiment to show adaptability, with results shown in the Appendix 9.

\section{Experiment}
\subsection{Datasets and Model Settings}
CoMET is pre-trained on mixed datasets, including HBN-EEG~\cite{shirazi2024hbn}, Stieger21~\cite{stieger2021continuous}, M3CV~\cite{huang2022m3cv} and SEED~\cite{zheng2015investigating}. For downstream tasks, ten widely used datasets are evaluated across multiple BCIs and clinical tasks. More details about the datasets are introduced in the Appendix 2.

\subsection{Pre-training}
For pre-training, We randomly divide 90\% of the training data into the training set and the remaining 10\% into the validation set. The mask ratio is set to 50\% according to CbraMod. We use the batchsize of 256 and the epochs of 100. The model is optimized using the AdamW optimizer. The max learning rate is set to 5e-4 and schedulered by CosineAnnealingLR. The pre-training experiment is conducted on a single node, equipped with 4 NVIDIA H100 (80GB) GPUs. The main software versions are Python 3.11.4, PyTorch 2.0.1 and CUDA 11.8. Subsequently, only simple and essential preprocessing steps are applied, including 0.5-70 Hz bandpass filtering, resampling to 200 Hz, 4 seconds segmenting without overlap, converting units to 0.1 mV. We use three configurations for CoMET with different depths of the encoder and hidden sizes: CoMET-Tiny (5M), CoMET-Base (19M), CoMET-Large (151M). More detailed settings please refer to the Appendix 3.

\subsection{Evaluation and Metrics}
We compare the CoMET with four SOTA brain foundation models (BIOT, LabraM, EEGPT and CBraMod). We re-implement baseline models using their official codes and released model weights. For data preprocessing, we follow the strategy of BIOT, EEGPT and CBraMod. All models use the same processed data on each dataset to ensure fairness. Downstream experiments were conducted on single node, equipped with 4 NVIDIA A100 (40GB) GPUs. Results shown in \textbf{Table \ref{tab:results1}} are obtained following the original baseline's downstream strategy, and additional evaluations of alternative adaptation strategies are provided in the Appendix 9.

\section{Results}

\begin{table*}[!ht]
\centering
\footnotesize
\setlength{\tabcolsep}{1mm}
\begin{tabular}{lccccccccccc}
\toprule
\multirow{2}{*}{\textbf{Model}} 
& \multicolumn{3}{c}{\textbf{BCIC IV 2A}} 
& \multicolumn{3}{c}{\textbf{BCIC IV 2B}} \\
\cmidrule(r){2-4} \cmidrule(r){5-7}
& \textbf{B. Acc} & \textbf{Kappa} & \textbf{F1} 
& \textbf{B. Acc} & \textbf{Kappa} & \textbf{F1} \\
\midrule
BIOT (NeurIPS'23)  &  41.44 $\pm$ 0.58$^{***}$ & 21.90 $\pm$ 1.19 & 37.37 $\pm$  0.56
& 67.78 $\pm$ 0.18$^{***}$ & 35.56 $\pm$ 0.36 & 66.41 $\pm$ 2.48 \\
LaBraM (ICLR'24)    & 52.49 $\pm$ 1.34$^{***}$ & 36.60 $\pm$ 2.12 & 52.32 $\pm$ 0.97 
& 71.39 $\pm$ 0.38$^{*}$ & 42.76 $\pm$ 0.76 & 70.68 $\pm$ 0.66 \\
EEGPT (NeurIPS'24)     & 51.37 $\pm$ 0.96$^{***}$ & 35.17 $\pm$ 1.26 & 49.73 $\pm$ 0.41 
& 67.33 $\pm$ 0.61$^{***}$ & 34.65 $\pm$ 1.23 & 67.35 $\pm$ 1.03 \\
CBraMod (ICLR'25)   & 55.85 $\pm$ 0.97$^{**}$ & 41.13 $\pm$ 1.30 & 55.08 $\pm$ 1.02 
& 67.35 $\pm$ 0.98$^{***}$ & 34.70 $\pm$ 1.96 & 71.06 $\pm$ 4.38 \\
CoMET-Tiny        & 58.53 $\pm$ 0.93 & 44.70 $\pm$ 0.50 & 58.00 $\pm$ 0.34 
& 71.42 $\pm$ 0.16 & 42.82 $\pm$ 0.31 & 70.48 $\pm$ 0.25 \\
CoMET-Base        & 61.66 $\pm$ 1.81 & 48.89 $\pm$ 2.42 & 60.52 $\pm$ 1.85 
& 72.71 $\pm$ 0.93 & 45.42 $\pm$ 1.86 & 72.52 $\pm$ 1.64 \\
CoMET-Large        & \textbf{62.75 $\pm$ 1.62} & \textbf{51.70 $\pm$ 1.84} & \textbf{63.37 $\pm$ 1.34} 
&\textbf{73.22 $\pm$ 0.81} & \textbf{46.32 $\pm$ 1.92} & \textbf{73.36 $\pm$ 1.74} \\

\toprule
\multirow{2}{*}{\textbf{Model}} 
& \multicolumn{3}{c}{\textbf{Large-5F}} 
& \multicolumn{3}{c}{\textbf{BCIC2020-3}} \\
\cmidrule(r){2-4} \cmidrule(r){5-7}
& \textbf{B. Acc} & \textbf{Kappa} & \textbf{F1} 
& \textbf{B. Acc} & \textbf{Kappa} & \textbf{F1} \\
\midrule
BIOT       & 32.16 $\pm$ 0.04$^{***}$ & 15.05 $\pm$ 0.05 & 31.02 $\pm$ 0.41 
& 31.31 $\pm$ 0.31$^{***}$ & 14.14 $\pm$ 0.39 & 30.66 $\pm$ 1.03 \\
LaBraM     & 31.41 $\pm$ 0.24$^{***}$ & 14.44 $\pm$ 0.27 & 31.21 $\pm$ 0.57 
& 36.10 $\pm$ 0.53$^{***}$ & 20.13 $\pm$ 0.66 & 35.30 $\pm$ 0.55 \\
EEGPT      & 31.48 $\pm$ 0.47$^{***}$ & 14.80 $\pm$ 0.62 & 29.78 $\pm$ 0.74 
& 37.05 $\pm$ 0.54$^{***}$ & 21.31 $\pm$ 0.68 & 36.22 $\pm$ 0.98 \\
CBraMod    & 34.60 $\pm$ 0.72$^{***}$ & 18.38 $\pm$ 0.88 & 33.86 $\pm$ 0.95 
& 47.29 $\pm$ 2.03$^{***}$ & 34.11 $\pm$ 2.54 & 46.66 $\pm$ 2.71 \\
CoMET-Tiny        & 34.66 $\pm$ 0.25 & 18.60 $\pm$ 0.33 & 33.75 $\pm$ 0.26 
& 52.97 $\pm$ 1.52 & 41.76 $\pm$ 2.19 & 52.80 $\pm$ 1.86 \\
CoMET-Base        & 37.03 $\pm$ 0.68 & 21.66 $\pm$ 0.82 & 35.93 $\pm$ 1.03 
& 54.43 $\pm$ 1.33 & 43.04 $\pm$ 1.66 & 54.12 $\pm$ 1.35 \\
CoMET-Large        & \textbf{38.97 $\pm$ 0.84} & \textbf{24.42 $\pm$ 0.94} & \textbf{39.24 $\pm$ 0.85} 
& \textbf{55.72 $\pm$ 1.42} & \textbf{44.84 $\pm$ 1.66} & \textbf{55.22 $\pm$ 1.03} \\

\toprule
\multirow{2}{*}{\textbf{Model}} 
& \multicolumn{3}{c}{\textbf{KaggleERN}} 
& \multicolumn{3}{c}{\textbf{FACED}} \\
\cmidrule(r){2-4} \cmidrule(r){5-7}
& \textbf{B. Acc} & \textbf{Kappa} & \textbf{F1} 
& \textbf{B. Acc} & \textbf{Kappa} & \textbf{F1} \\
\midrule
BIOT       & 50.64 $\pm$ 0.01$^{***}$ & 1.62 $\pm$ 0.03 & 71.53 $\pm$ 0.01 
& 33.77 $\pm$ 0.39$^{***}$ & 25.32 $\pm$ 0.46 & 33.32 $\pm$ 0.3\\
LaBraM     & 51.95 $\pm$ 0.81$^{***}$ & 4.38 $\pm$ 1.75 & 77.21 $\pm$ 1.82
&  35.13 $\pm$ 1.76$^{**}$ &  26.87 $\pm$ 2.00 & 35.05 $\pm$ 1.9\\
EEGPT      & 54.92 $\pm$ 0.04$^{***}$ & 11.89 $\pm$ 0.12 & 76.77 $\pm$ 0.04
&  31.56 $\pm$ 0.11$^{***}$ & 22.93 $\pm$ 0.14 & 31.35 $\pm$ 0.19\\
CBraMod    & 53.92 $\pm$ 0.02$^{***}$ & 8.15 $\pm$ 0.05 & 76.49 $\pm$ 0.04
& 34.35 $\pm$ 4.34$^{***}$ & 25.83 $\pm$ 4.86 & 33.51 $\pm$ 5.1\\
CoMET-Tiny        & 55.45 $\pm$ 0.12 & 13.28 $\pm$ 0.41 & 77.74 $\pm$ 0.78
& 33.28 $\pm$ 0.31 &  24.78 $\pm$ 0.36 & 33.12 $\pm$ 0.35\\
CoMET-Base        & 57.08 $\pm$ 0.12 & 14.63 $\pm$ 0.33 & 78.92 $\pm$ 0.53
&  38.40 $\pm$ 0.55 & 30.49 $\pm$ 0.64 & 38.02 $\pm$ 0.75\\
CoMET-Large        & \textbf{58.78 $\pm$ 0.88} & \textbf{15.24 $\pm$ 0.92} & \textbf{79.66 $\pm$ 1.43}
& \textbf{39.02 $\pm$ 0.38} & \textbf{31.47 $\pm$ 0.47}& \textbf{39.13 $\pm$ 0.59} \\

\toprule
\multirow{2}{*}{\textbf{Model}} 
& \multicolumn{3}{c}{\textbf{THUBenchmark}} 
& \multicolumn{3}{c}{\textbf{PhysioP300}} \\
\cmidrule(r){2-4} \cmidrule(r){5-7}
& \textbf{B. Acc} & \textbf{Kappa} & \textbf{F1} 
& \textbf{B. Acc} & \textbf{Kappa} & \textbf{F1} \\
\midrule
BIOT       & 74.17 $\pm$ 0.06$^{***}$ & 73.50 $\pm$ 0.07 & 74.24 $\pm$ 0.05 
& 50.04 $\pm$ 0.12$^{***}$ & 0.09 $\pm$ 0.25 & 6.31 $\pm$ 0.88 \\
LaBraM     & 91.43 $\pm$ 0.13 & 91.21 $\pm$ 0.13 & 91.43 $\pm$ 0.13
&  54.26 $\pm$ 0.22$^{***}$ & 8.38 $\pm$ 0.43 & 56.86 $\pm$ 1.38\\
EEGPT      & 82.59 $\pm$ 0.09$^{***}$ & 82.15 $\pm$ 0.09 & 82.57 $\pm$ 0.08 
& 61.07 $\pm$ 0.27 & 22.15 $\pm$ 0.52 & 54.24 $\pm$ 0.48 \\
CBraMod    & 91.45 $\pm$ 0.24 & 91.23 $\pm$ 0.25 & 91.43 $\pm$ 0.24 
& 59.46 $\pm$ 0.01$^{***}$ & 18.85 $\pm$ 0.03 & 55.68 $\pm$ 0.03 \\
CoMET-Tiny & 91.72 $\pm$ 1.85 & 90.17 $\pm$ 0.67 & 89.66 $\pm$ 1.49 
& 61.36 $\pm$ 0.50 & 22.58 $\pm$ 1.00 & 63.44 $\pm$ 0.58 \\
CoMET-Base & 92.01 $\pm$ 1.09 & 91.81 $\pm$ 1.10 & 92.00 $\pm$ 1.11 
& 62.36 $\pm$ 0.71 & 24.55 $\pm$ 1.41 & 60.52 $\pm$ 3.63 \\
CoMET-Large & \textbf{92.74 $\pm$ 1.61} & \textbf{92.65 $\pm$ 2.46} & \textbf{93.50 $\pm$ 0.81} 
& \textbf{63.86 $\pm$ 1.78} & \textbf{25.05 $\pm$ 2.01} & \textbf{61.72 $\pm$ 2.28} \\

\toprule
\multirow{2}{*}{\textbf{Model}}
& \multicolumn{3}{c}{\textbf{TUAB}}
& \multicolumn{3}{c}{\textbf{TUEV}}\\
\cmidrule(r){2-4}\cmidrule(r){5-7}
& \textbf{B.\,Acc} & \textbf{AUC} & \textbf{--}
& \textbf{B.\,Acc} & \textbf{Kappa} & \textbf{F1}\\
\midrule
BIOT       &  79.59 $\pm$ 0.57$^{**}$ & 88.15 $\pm$ 0.43 & --
&  52.81 $\pm$ 2.25$^{**}$ & 52.73 $\pm$ 2.49 & 74.92 $\pm$ 0.82\\
LaBraM$^\triangle$     & 81.40 $\pm$ 0.19 & 90.22 $\pm$ 0.09 & --
&  64.09 $\pm$ 0.65 & 66.37 $\pm$ 0.93 & 83.12 $\pm$ 0.52\\
EEGPT      &  79.83 $\pm$ 0.30$^{*}$ & 87.18 $\pm$ 0.50 & --
& 62.32 $\pm$ 1.14$^{*}$ & 63.51 $\pm$ 1.34 & 81.87 $\pm$ 0.63\\
CBraMod$^\triangle$    & \textbf{82.29 $\pm$ 0.22} & \textbf{92.27 $\pm$ 0.11} & --
&  \textbf{66.71 $\pm$ 1.07} & \textbf{67.72 $\pm$ 0.96} & \textbf{83.42 $\pm$ 0.64}\\
CoMET-Tiny & 80.02 $\pm$ 1.34 & 87.58 $\pm$ 1.08 & --
&  60.31 $\pm$ 2.40 & 62.83 $\pm$ 0.84 & 80.32 $\pm$ 0.85\\
CoMET-Base & 81.87 $\pm$ 0.19 & 89.16 $\pm$ 0.24 & --
&  62.31 $\pm$ 2.40 & 64.05 $\pm$ 1.51 & 80.47 $\pm$ 2.11\\
CoMET-Large & 82.02 $\pm$ 1.05 & 91.04 $\pm$ 0.97 & --
&  62.97 $\pm$ 1.31 & 66.95 $\pm$ 0.31 & 82.88 $\pm$ 1.93\\
\bottomrule

\end{tabular}
\caption{Performance comparison of models on ten downstream datasets with Wilcoxon Signed-Rank Test (CoMET-Large vs. others) ($^{*}$:p$<$0.05, $^{**}$:p$<$0.01, $^{***}$:p$<$0.001).}
\label{tab:results1}
\end{table*}

\begin{figure*}[!t]
\centering
\includegraphics[width=0.9\textwidth]{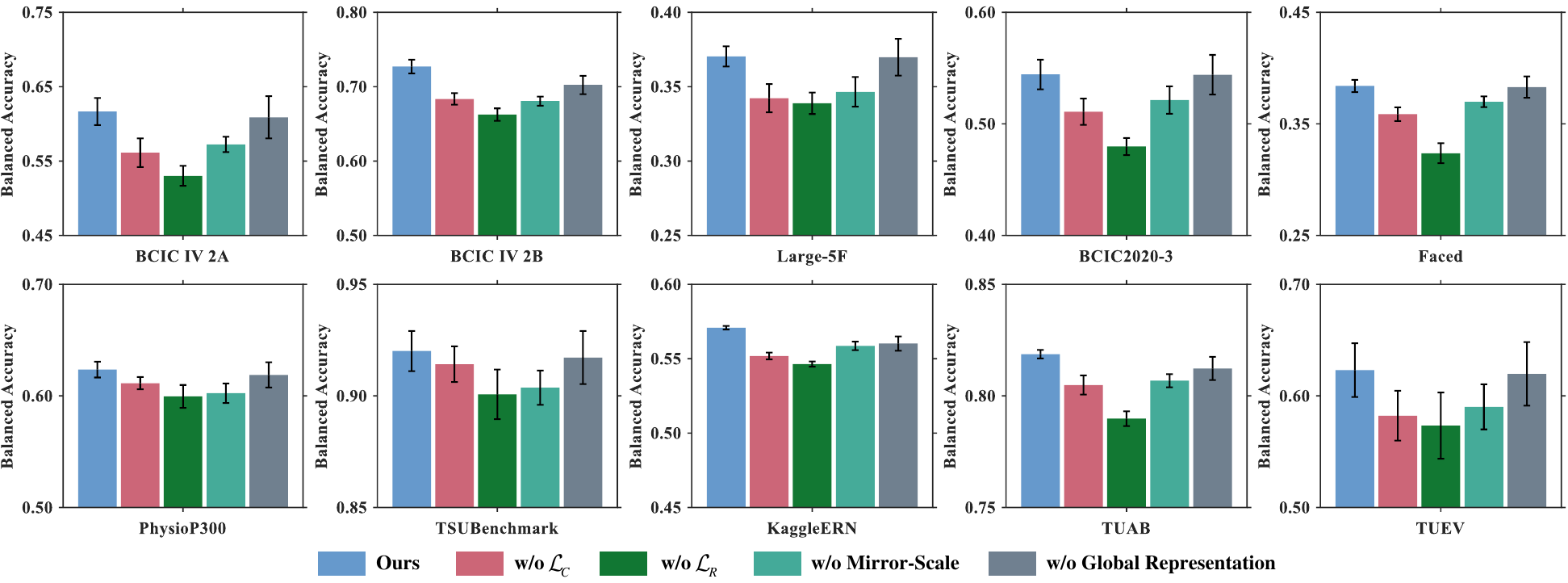}
\vspace{-4pt}
\caption{Ablation Study. We study the effort of two losses (w/o  $\mathcal{L}_R$ and w/o $\mathcal{L}_C$), augmentation method (random masking w/o Mirror-Scale) and contrastive feature extraction module (w/o Global Representation) on CoMET-Base.}
\label{ablation}
\vspace{-10pt}
\end{figure*}

We conducted a comprehensive evaluation of our proposed CoMET models against strong baselines across diverse EEG tasks. The full results are summarised in \textbf{Table \ref{tab:results1}}. 
On the widely used motor imagery datasets BCIC IV 2A, CoMET-Large attains balanced accuracy of 62.75\% ± 1.62 on 2A, exceeding the previous SOTA CBraMod (55.85\% ± 0.97) by an absolute margin of 6.9\%. Similarly, on BCIC IV 2B, CoMET-Large reaches 73.22\%±0.81, outperforming the strongest baseline LaBraM (71.39\% ± 0.28) by 1.83\%.
For Large-5F (five-finger imagery), CoMET-Large achieves balanced accuracy of 38.97\% ± 0.84, surpassing the best baseline by 4.37\%. On the BCIC-2020-3 (imagined speech), CoMET-Large achieves the highest performance of 55.72\% ± 1.42, outperforming the baseline by 5.81\%. For emotion recognition dataset KaggleERN and FACED, CoMET-Large surpasses the leading baseline by 2.67\% and 3.14\% in balanced accuracy, respectively. As for BCI tasks that required visual stimuli (THUBenchmark, and PhysioP300), CoMET-Large also achieved the best banlanced accuracies of 92.74\% and 63.86\%, respectively, which are 1.29\% and 2.79\% higher than the best baseline CBraMod and EEGPT. 

On TUAB and TUEV, CoMET-Large marginally underperforms LabraM and CBraMod. However, it is important to note that the TUH dataset is used by CBraMod in pre-training, which includes both TUAB and TUEV. LabraM employs TUEG from TUH for pre-training, which shares a similar distribution with TUAB and TUEV. None of the pre‑training data of CoMET originates from the TUH family (TUH, TUAB, TUEV, etc.). When the comparison is restricted to models pre‑trained on different source datasets, CoMET still achieves the best performance. Moreover, CoMET's performance increases monotonically from Tiny to Base, and Large, adhering to the scaling law and indicating headroom for further improvements as model capacity grows.

\subsection{Ablation Study}
In order to evaluate the effectiveness of key components in CoMET, we conduct an ablation study in the pre-training stage as follows: 1) w/o $\mathcal{L}_{C}$: only use masked reconstruction in pre-training; 2) w/o $\mathcal{L}_R$: only use contrastive learning in pre-training; 3) w/o mirror-scale augmentation: use random cropped masking to construct contrastive view augmentation; 4) w/o global representation: Append linear projection after encoder to get global features.
The ablation results in \textbf{Figure \ref{ablation}} demonstrate that each component of CoMET is indispensable and mutually reinforcing: MEM serves as the backbone, guaranteeing the model's ability to capture local, shallow features, whereas CL further reinforces its capacity for global modelling. Removing either the $\mathcal{L}_R$ or the $\mathcal{L}_C$—instead of employing both concurrently—reduces downstream classification accuracy by up to 6.98\% (TUEV) and 5.54\% (BCIC IV 2A), demonstrating that MEM and CL emphasize different aspects during pre-training yet complement one another effectively. $\mathcal{L}_C$ contributes less on THUBenchmark, possibly because SSVEP focus more on the frequency domain features of specific brain regions~\cite{lin2006frequency}. Mirror-Scale Augmentation ensures that CL aligns global invariant features across different views of the same sample; when the augmented views are instead generated by random masking, the CL branch contributes little and can even be detrimental. The global token contributes little improvement on mean accuracy, but it markedly reduces variance, thereby enhancing the stability of the pre-trained model.

\section{Discussion}
\begin{figure}[!t]
\centering
\includegraphics[width=0.38\textwidth]{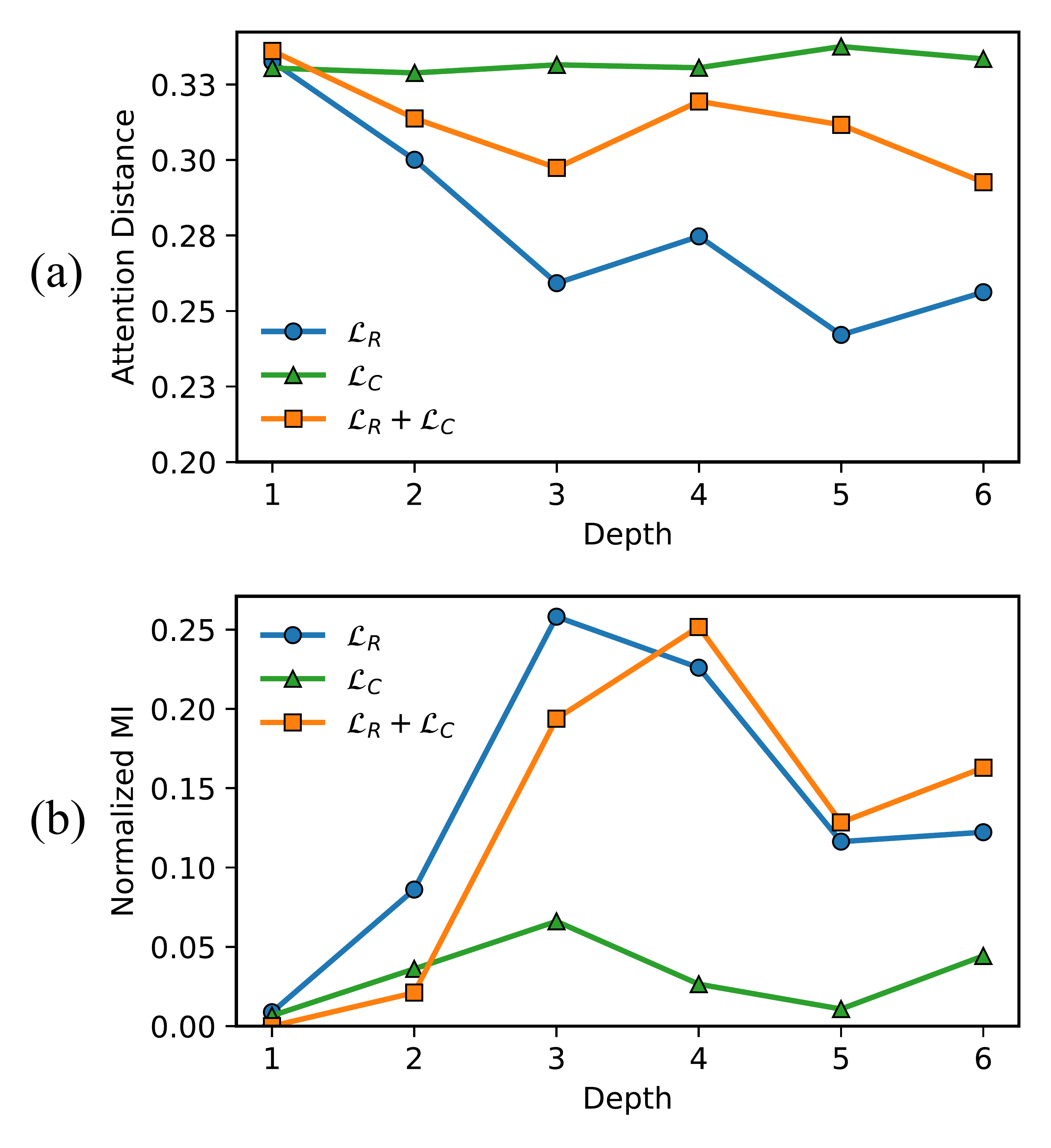}
\vspace{-9pt}
\caption{Analysis of (a) attention distance and (b) normalized mutual information across layers. Larger attention distance indicates wider effective receptive fields, and lower Normalized MI indicates attention homogeneous collapse.}
\vspace{-10pt}
\label{attention_distance}
\end{figure}

To understand how the reconstructive loss ($\mathcal{L}_R$) and contrastive loss ($\mathcal{L}_C$) contribute to the model's receptive fields, we analyze the average distance between the query token weights and key token weights of different channels~\cite{dosovitskiyimage}. Channels use relative coordinates normalized to (0, 1). Figure \ref{attention_distance} (a) shows $\mathcal{L}_C$ captures wider global information and stabilizes between layers. While $\mathcal{L}_R$ focus on local channel's relationships, $\mathcal{L}_R + \mathcal{L}_C$ (ours) can effectively increase the receptive fields. We also use normalized mutual information (NMI) ~\cite{strehl2002cluster} to measure the attention collapse. NMI computes the mutual information between query and key tokens $I(q,k)$ from different channels, and normalize it by their marginal entropies: $\frac{I(q,k)}{\sqrt{H(q)H(k)}}$. Lower NMI indicates attention map are less dependent on query tokens and attention collapse into homogeneity. As shown in Figure \ref{attention_distance} (b), $\mathcal{L}_C$ exhibits homogeneous attention collapse across all layers, whereas $\mathcal{L}_R + \mathcal{L}_C$ (ours) markedly enhances attention diversity, even outperforming $\mathcal{L}_R$ in final three layers. This analysis visually confirms the motivation of our article: MEM and CL focus on different EEG features and they are able to gain their respective advantages non-destructively through aggregation.

\section{Conclusion}
In this study, we propose a contrastive-masked EEG self-supervised framework, training the 151 million parameter brain foundation model CoMET. We employ the masked autoencoder with redesigned patching and embedding strategies for EEG as the backbone, and devise a novel contrastive learning framework with mirror-scale augmentation to effectively to merge the local fast convergence ability of masked EEG reconstruction and the global discrimination ability of contrastive learning. Experiments on ten downstream datasets demonstrate CoMET’s SOTA capacity to extract both local context–sensitive and globally discriminative EEG representations, outperforming prior foundation models based solely on contrastive learning or masked reconstruction, especially on these tasks that relies on connection between multi brain regions. The proposed foundation model CoMET has been designed with full consideration of the characteristics of EEG signals, and the new global modeling idea provides a meaningful contribution to the study of brain foundation models, promoting the AI utilization in real-world BCIs, clinical, and healthcare applications.

\bibliography{main}

\clearpage

\section{Appendix}
\begin{table*}[!t]
\centering
\begin{tabular}{cccccccc}
\hline
  & \textbf{Datasets}     & \textbf{Task}                & \textbf{Subjects} & \textbf{Channel} & \textbf{Duration} & \textbf{Samples} & \textbf{Classes} \\ \hline
\multirow{4}{*}{\textbf{Pre-training Datasets}} & HBN          & Multi-Tasks         & 3155     & 128     & 4        &  1497026       & -       \\
  & Stieger21    & Motor Imagery       & 62       & 64      & 4        & 269099        & -       \\
  & M3CV         & Multi-Tasks         & 106      & 64      & 4        & 116863        & -       \\
  & SEED       & Emotion Recognition & 15       & 62      & 4        &   30375      & -       \\ \hline
\multirow{10}{*}{\textbf{Downstream Datasets}} & BCIC IV 2A   & MI                  & 9        & 22      &    4      & 5184    & 4       \\
  & BCIC IV 2B   & MI                  & 9        & 3       &     4     & 900     & 2       \\
  & PhysioP300   & P300                & 10       & 64      &    2      & 2532    & 2       \\
  & KaggleERN    & ERP                 & 26       & 56      &    2      &    4448     & 2       \\
  & FACED        & Emotion Recognition & 123      & 32      &    10      &   10332      & 9       \\
  & THUBenchmark & SSVEP               & 35       & 64      & 2        & 8400    & 40      \\
  & Large-5F     & Finger MI           & 13       & 38      &    4.5      &    18071     & 5       \\
  & BCIC2020-3   & Speech Imagery     & 20       & 62      &     3     & 8000    & 5       \\
  & TUAB         & Clinical            & 2383     & 23      &    10      &     409083    & 2       \\
  & TUEV         & Artifact Detection  & 288      & 21      &      5    &    112237     & 6       \\ \hline
\end{tabular}
\caption{Detailed information of the pre-training and downstream datasets}
\label{table1}
\end{table*}
\subsection{1. Pre-training Datasets}
\begin{enumerate}
    \item HBN-EEG dataset \cite{shirazi2024hbn} provides high-density (128-channel) EEG from more than 3 000 subjects collected as part of the Healthy Brain Network project. Participants completed six paradigms covering passive conditions—resting state, visual surround-suppression and movie watching—and active tasks of contrast-change detection, sequence learning and symbol search. Recordings were made at 500 Hz (0.1-100 Hz online band-pass, 60 Hz notch) and Cz reference. For further preprocessing we applied band-pass filter 0.5-70 Hz, segment the continuous data into non-overlapping 4 seconds windows with 1 second intervel and down-sample to 200 Hz. Following three pre-training datasets are applied similar preprocessing as HBN-EEG.
    \item Stieger21 dataset~\cite{stieger2021continuous} contains longitudinal EEG from 62 healthy adults who practiced online sensorimotor-rhythm BCI control over 7-11 sessions (598 sessions, \textgreater600 h, 269 000 trials). Each session comprised four continuous cursor-control tasks: horizontal (left/right MI), vertical (up/down MI), two-dimensional (combined MI) and rest calibration, with real-time feedback every 40 ms. Signals were recorded with a 64-channel BioSemi cap (10–5 montage) at 1 kHz and hardware-filtered 0.1–200 Hz with a 60 Hz notch.
    \item M3CV dataset~\cite{huang2022m3cv} is a large-scale “multi-subject, multi-session, multi-task” resource designed for biometric and variability studies. EEG was recorded from 106 subjects (95 returned for a second visit) while they performed six broad paradigms—resting-state, transient sensory, steady-state sensory, cognitive oddball, motor execution and selective-attention SSVEP—yielding 14 concrete task types and 120 000 labeled epochs. Signals were acquired with 64-channel EasyCap nets at 1 000 Hz (0.1–100 Hz pass-band, 50 Hz notch) with FCz reference and AFz ground.
    \item SEED dataset~\cite{zheng2015investigating} comprises emotion-elicitation EEG from 15 university students who viewed fifteen 4-min film clips intended to provoke positive, neutral or negative affect in three separate sessions spaced one week apart. EEG (62 channels, NeuroScan cap, 1 000 Hz, 0.05–100 Hz online band-pass, 50 Hz notch) and synchronous eye-tracking were recorded; electrodes were referenced to CPz.
\end{enumerate}
\subsection{2. Downstream Datasets}
\begin{enumerate}
    \item BCIC-IV-2A dataset~\cite{bciciv2a} comprises EEG recordings from nine subjects performing four-class motor imagery (MI) tasks: left hand (Class 1), right hand (Class 2), both feet (Class 3), and tongue (Class 4). Each subject participated in two sessions conducted on separate days, each containing six runs and a total of 288 trials. EEG signals were acquired using 22 Ag/AgCl electrodes arranged according to the international 10–20 system, referenced to the left mastoid, sampled at 250 Hz, and originally bandpass-filtered between 0.5 Hz and 100 Hz with a 50 Hz notch filter to suppress line noise. Additionally, three monopolar EOG channels were recorded to facilitate ocular artifact removal but are not utilised in classification. For preprocessing, we apply a 0.5–38 Hz bandpass filter to retain relevant sensorimotor rhythms while discarding high-frequency noise and slow drifts. Each trial segment is extracted over a 4-second window post cue onset and subsequently downsampled to 200 Hz to reduce computational overhead. We retain all 22 EEG channels for downstream processing. To ensure subject-independent evaluation, we adopt a leave-one-subject-out (LOSO) cross-validation strategy throughout all experiments.
    \item BCIC-IV-2B dataset \cite{bciciv2b} involves binary motor imagery tasks (left hand vs. right hand) performed by nine right-handed subjects. Each subject participated in two initial screening sessions without feedback (120 trials per session), followed by three feedback sessions using a smiley-face interface (80 trials per session). EEG data were recorded from three bipolar electrode channels (C3, Cz, and C4), sampled at 250 Hz, and filtered with a 0.5–100 Hz bandpass filter alongside a 50 Hz notch filter to suppress power line interference. For preprocessing, we apply a bandpass filter in the range of 0.5–38 Hz to retain key motor-related rhythms while attenuating noise. Each trial is segmented into a 4-second epoch post-cue and downsampled to 200 Hz. All three bipolar channels are retained for analysis. We employ a LOSO cross-validation strategy to ensure subject-independent evaluation.
    \item PhysioP300 dataset \cite{physiop300} consists of EEG recordings from nine subjects (8, 10, and 12 are removed for a fair comparison with the BENDR and EEGPT methdos) performing a visual P300 speller task based on the classic row-column paradigm. Participants were instructed to attend to specific target characters in a 6x6 matrix, where random sequences of row and column flashes served as stimuli to elicit P300 event-related potentials (ERPs) in response to target cues. The original recordings were acquired using 64-channel EEG caps following the international 10-20 system, downsampled to 250 Hz, and bandpass-filtered between 0.15 Hz and 5 Hz. For our experiments, we retain 58 EEG channels, apply a 120 Hz low-pass filter to reduce high-frequency noise, and resample the data to 200 Hz. Stimulus-locked epochs of 2 seconds duration (-0.7 s to +1.3 s relative to stimulus onset) are extracted for each flash event. We frame the task as a binary classification problem (target vs. non-target) and adopt a LOSO cross-validation scheme for subject-independent evaluation.
    \item KaggleERN dataset \cite{kaggleern}, which contains EEG recordings from 26 healthy participants engaged in a P300-based speller paradigm augmented with online error detection and correction. Subjects were instructed to focus on target letters within a 6x6 flashing matrix and engaged in two spelling modes: fast mode (2 sequences per trial) and slow mode (4 sequences per trial). The task was designed to elicit error-related potentials (ErrPs) following incorrect classifier outputs, enabling automatic correction by selecting the classifier's second-best prediction when an error was detected. For our experiments, EEG trials were resampled to 200 Hz, and binary classification (correct vs. incorrect feedback) was performed. We adopted a 4-fold cross-subject validation strategy, where each fold utilized data from 12 subjects for training and the remaining 10 subjects for testing. A total of 19 channels were used in the analysis, and data within the time window of -0.7 s to +1.3 s relative to stimulus onset were cropped and used for classification.
    \item FACED dataset \cite{faced} consists of 32-channel EEG recordings from 123 subjects who viewed 28 video clips designed to induce nine distinct emotional categories: four negative emotions (anger, fear, disgust, sadness), four positive emotions (amusement, inspiration, joy, tenderness), and a neutral condition. The data were recorded at 250 Hz or 1000 Hz using the 10-20 system. Official preprocessing steps involved bandpass filtering between 0.05 and 47 Hz, independent component analysis (ICA) for artifact removal, and re-referencing to a common average reference, resulting in 30 clean channels. For our experiments, the data were segmented into 10-second clips, resampled to 200 Hz, and evaluated using a 4-fold cross-subject validation strategy, where each fold used two-thirds of the subjects for training and the remaining one-third for testing. We used 30 channels from the officially re-referenced dataset for classification.
    \item THUBenchmark dataset \cite{benchmark} consists of EEG recordings from 35 subjects engaged in a 40-class steady-state visual evoked potential (SSVEP) task. Participants focused on target characters modulated by distinct frequency/phase combinations to elicit class-specific SSVEP responses. EEG signals were recorded using 64 electrodes at a sampling rate of 1000 Hz. For our experiments, we extracted 2-second stimulation windows from each trial, excluding the initial 140 ms visual latency. The signals were bandpass filtered between 0.5 Hz and 45 Hz, downsampled to 200 Hz, and nine occipital-parietal channels were selected for analysis. Evaluation was conducted using 4-fold cross-subject validation, with each fold comprising 80\% of the subjects for training and the remaining 20\% for testing.
    \item Large-5F dataset \cite{finger} comprises EEG recordings from nine subjects performing five-class motor imagery tasks involving individual finger movements: thumb, index, middle, ring, and pinkie. Data were collected as part of a multi-paradigm EEG study using 22 electrodes placed according to the international 10–20 system, with recordings sampled at either 200 Hz or 1000 Hz. A bandpass filter of 0.53–70 Hz (or 100 Hz) and a 50 Hz notch filter were applied during acquisition. Each subject completed up to 75 sessions, resulting in over 60,000 trials across all paradigms. For our experiments, we focused on the five-finger (5F) motor imagery paradigm. Cue-aligned EEG segments were extracted and temporally interpolated to a uniform maximum duration of 4.5 seconds to ensure consistency across trials. The signals were then downsampled to 200 Hz and bandpass filtered between 0.5 Hz and 45 Hz to remove low-frequency drifts and high-frequency noise. All 22 EEG channels were retained, and evaluation was performed using a leave-one-subject-out (LOSO) cross-validation strategy to assess subject-independent performance.
    \item BCIC2020-3 dataset \cite{speech} comprises EEG recordings from 15 healthy subjects aged 20 to 30 years, performing imagined speech tasks involving five phrases: "hello," "help me," "stop," "thank you," and "yes." EEG signals were acquired using 64 electrodes arranged according to the international 10-20 system, with ground electrode at Fpz and reference at FCz. Impedances were maintained below 15 k$\Omega$ to ensure signal quality. Each trial consisted of a 2-second auditory cue followed by a 2-second imagery phase. Each phrase class included 70 trials, with 60 trials designated for training and 10 for validation. For experimental evaluation, 3-second epochs were extracted by concatenating the 1-second cue period and the 2-second imagery phase. The data were downsampled to 200 Hz to reduce computational load. All 64 EEG channels were retained. We employed a cross-session validation scheme to assess the model's ability to generalize across recording sessions.
    \item TUAB dataset \cite{tuh} facilitates binary classification of normal versus abnormal adult EEG recordings based on background brain activity. EEG signals were recorded using 64 electrodes placed according to the international 10-20 system, with ground at Fpz and reference at FCz, maintaining impedances below 15 k$\Omega$. The raw data were sampled at either 200 Hz or 1000 Hz and bandpass filtered between 0.53 Hz and 70 Hz (or 100 Hz) alongside a 50 Hz notch filter to reduce line noise. For our experiments, we selected 23 representative channels, downsampled the signals to 200 Hz, and extracted 10-second segments. Evaluation was performed using 4-fold cross-subject validation to ensure robust generalisation across patients.
    \item TUEV dataset \cite{tuh}, a subset of the Temple University Hospital EEG Corpus, supports classification of six clinically relevant EEG event types: spike and sharp waves (SPSW), periodic lateralized epileptiform discharges (PLED), generalized periodic epileptiform discharges (GPED), artifacts (ARTF), eye movements (EYEM), and background activity (BCKG). The dataset comprises 16,986 sessions from 10,874 subjects, recorded using a variable number of EEG channels (typically 31) at sampling rates of 250, 256, 400, or 512 Hz. For our experiments, we selected 23 standard EEG channels, downsampled the signals to 200 Hz, and segmented the data into 5-second epochs. A 4-fold cross-subject validation scheme was adopted to evaluate model performance across subjects.
\end{enumerate}

\subsection{3. More Details For Experimental Settings}
\textbf{Data preprocessing}: For \textbf{pre-training}, we applied minimum necessary data preprocessing. We first apply a 0.5 HZ to 70 HZ band-pass filter on EEG signals and resample them to 200 HZ. Then we normalize the EEG signals by setting the unit to 0.1 mV to guarantee the value mainly between -1 to 1 \cite{jianglarge}. EEG signals are segmented into 4 seconds non-overlap samples according to LaBraM \cite{jianglarge} and EEGPT \cite{wang2024eegpt}. During pre-training, we select 62 EEG channels that are commonly shared across all pre-training datasets. Although CoMET supports varying channel configurations in downstream tasks, using unbalancing number of channels during pre-training can lead to inbalancing channel embeddings, ultimately degrading model performance. In contrast, 62 channel embeddings cover the majority of channel configurations in EEG experiment, and ensure consistent representation for different channels. For \textbf{downstream tasks}, we adopt dataset-specific preprocessing strategies to account for structural differences across datasets. Detailed procedures are provided in the dataset description section. In summary, downstream samples vary in terms of filtering frequency bands, sample duration, and the number of EEG channels.

To ensure the reproducibility of our experimental results, we provide more details about the hyperparameters settings on CoMET pre-training in Table \ref{table-pretraining} and downstream tasks in Table \ref{table-downstream-settings}.

\begin{table*}[h]
\centering
\begin{tabular}{ccccc}
\hline
\multicolumn{2}{c}{Hyperparameters}                            & CoMET-Tiny & CoMET-Base & CoMET-Large \\ \hline
\multirow{4}{*}{Patching Embedding}  & Input dimension         & 1          & 1          & 1           \\
                                     & Output dimension        & 256        & 512        & 1024         \\
                                     & Kernel size             & \multicolumn{3}{c}{(1, 50)}           \\
                                     & Stride                  & \multicolumn{3}{c}{(1, 50)}           \\ \hline
\multirow{4}{*}{Transformer Encoder} & Layers                  & 6          & 6          & 12           \\
                                     & Hidden dimension        & 256        & 512        & 1024         \\
                                     & Heads                   & 4          & 8          & 16          \\
                                     & Feed-forward dimension  & 1024       & 2048       & 4096        \\ \hline
\multirow{4}{*}{Transformer Decoder} & Layers                  & 2          & 4          & 6           \\
                                     & Hidden dimension        & 384        & 384        & 384         \\
                                     & Heads                   & 4          & 8          & 16          \\
                                     & Feed-forward dimension  & 1536       & 1536       & 1536        \\ \hline
\multirow{9}{*}{Pre-training}         & Epochs                  & \multicolumn{3}{c}{100}               \\
                                     & Batchsize               & \multicolumn{3}{c}{256}               \\
                                     & Peak learning rate      & \multicolumn{3}{c}{5e-4}              \\
                                     & Learning rate scheduler & \multicolumn{3}{c}{CosineAnnealingLR} \\
                                     & Optimizer               & \multicolumn{3}{c}{AdamW}             \\
                                     & Adam $\beta$              & \multicolumn{3}{c}{(0.9 0.999)}       \\
                                     & Weight decay scheduler            & \multicolumn{3}{c}{CosineWDSchedule (1e-6)}              \\
                                     & Momentum scheduler      & \multicolumn{3}{c}{(0.996, 1)}        \\
                                     & Temperature      & \multicolumn{3}{c}{0.1}        \\
                                     & Mask ratio              & \multicolumn{3}{c}{0.5}               \\ \hline
\end{tabular}
\caption{Hyperparameters for CoMET Pre-training}
\label{table-pretraining}
\end{table*}

\begin{table*}[h]
\centering
\begin{tabular}{ccccc}
\hline
\multicolumn{2}{c}{Hyperparameters}                            & CoMET-Tiny & CoMET-Base & CoMET-Large \\ \hline
\multirow{2}{*}{Linear probing}      & Input dimension         & {256, patch} & {512, patch} & {1024, patch} \\
                                     & Output dimension        & {16, classes}            & {32, classes}            & {48, classes}       \\ \hline
                                     & Epochs                  & \multicolumn{3}{c}{100}                                                        \\
                                     & Batchsize               & \multicolumn{3}{c}{64}                                                         \\
                                     & Peak learning rate      & \multicolumn{3}{c}{1e-3}                                                       \\
                                     & Optimizer               & \multicolumn{3}{c}{AdamW}                                                      \\
                                     & Adam $\beta$              & \multicolumn{3}{c}{(0.9 0.999)}       \\ \hline                 
\end{tabular}
\caption{Hyperparameters for CoMET Downstream}
\label{table-downstream-settings}
\end{table*}

\subsection{4. Pre-training visualization}
\begin{figure}[!ht]
\centering
\includegraphics[width=0.45\textwidth]{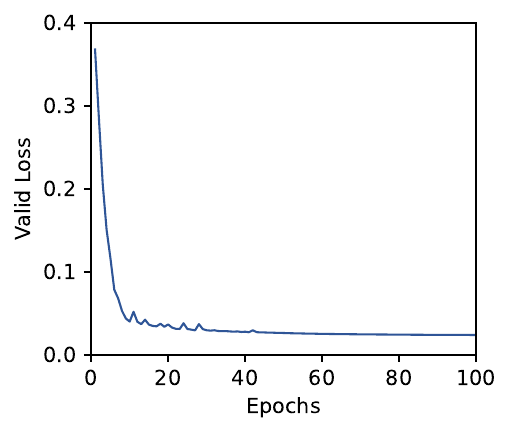}
\caption{The reconstruction loss ($\mathcal{L}_R$) of pre-training}
\label{fig6}
\end{figure}

\begin{figure}[!ht]
\centering
\includegraphics[width=0.45\textwidth]{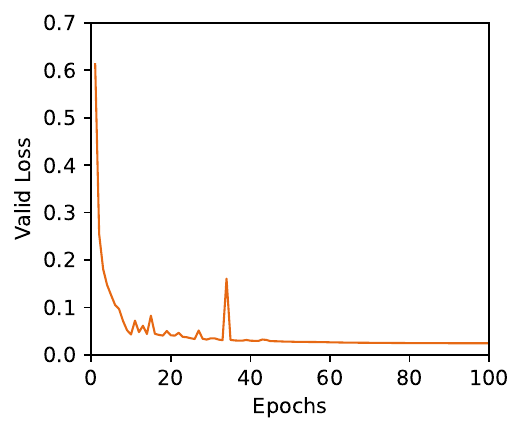}
\caption{The contrastive loss ($\mathcal{L}_C$) of pre-training}
\label{fig7}
\end{figure}
In Figure \ref{fig6} and Figure \ref{fig7}, we present the reconstruction loss ($\mathcal{L}_R$) curve and contrastive loss ($\mathcal{L}_C$) curve of pre-training. It is evident that both losses decrease rapidly between epochs 1 and 20. When the epoch exceeds 20, both losses tend to stabilize, with the contrast loss exhibiting fluctuations at epoch 35. Throughout pre-training, the two loss functions exhibited no reciprocal constraint, The common downward trajectory indicates that CoMET can effectively acquires local EEG representations through the reconstruction task and global discriminative representations through the contrastive task.
\subsection{5. Scaling laws}
\begin{figure*}[!htbp]
\centering
\includegraphics[width=0.9\textwidth]{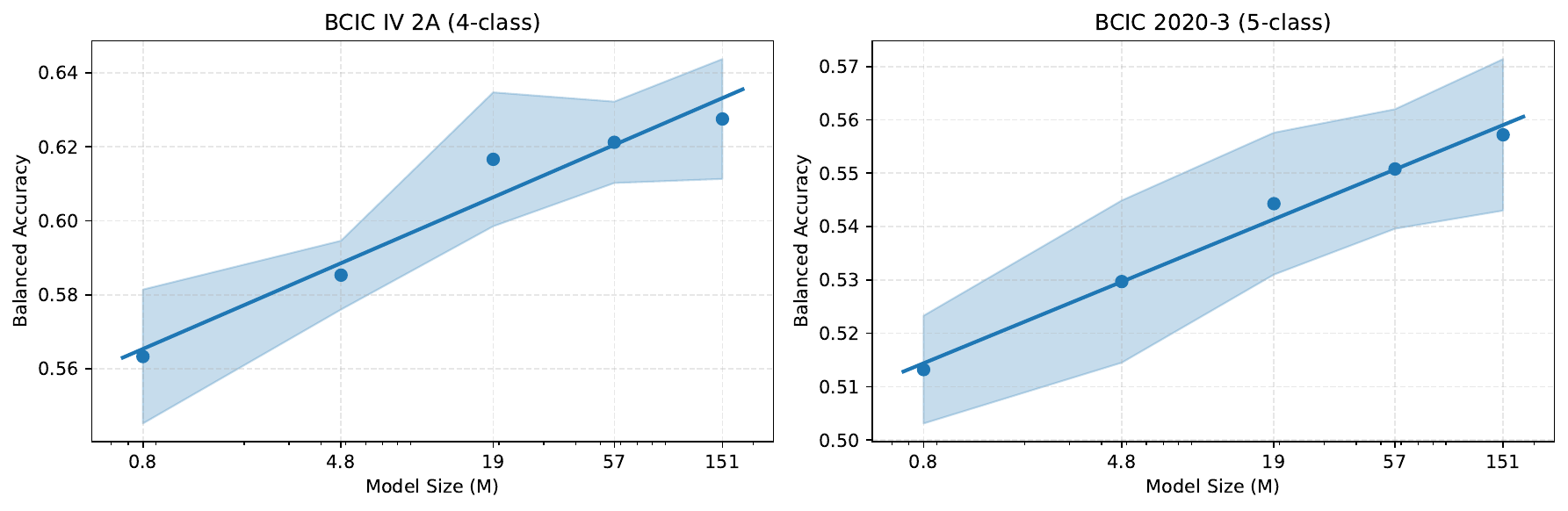}
\caption{The scaling law with model size ($M$) and balanced accuracy. BCIC IV 2A: $BAC = 0.013 \cdot ln(M) + 0.568, (R^{2} = 0.949)$; BCIC2020-3: $BAC = 0.009 \cdot ln(M) + 0.516, (R^{2} = 0.989)$}
\label{fig-scaling-model}
\end{figure*}

\begin{figure*}[!htbp]
\centering
\includegraphics[width=0.9\textwidth]{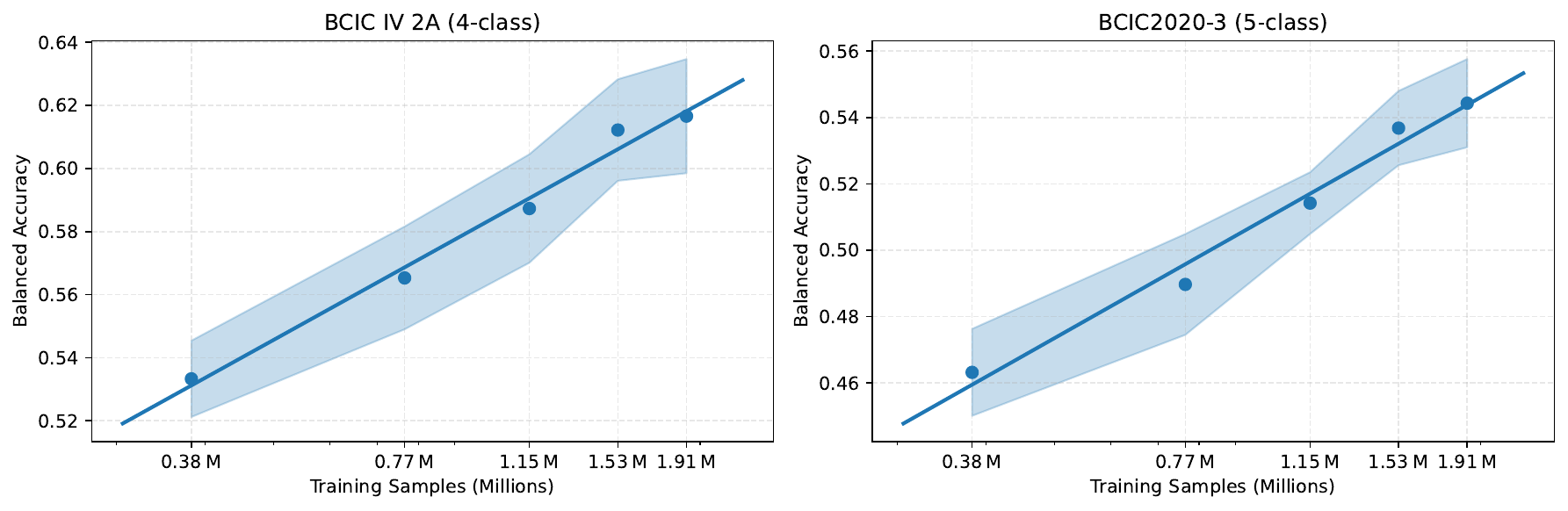}
\caption{The scaling law with training data size ($N$) and balanced accuracy. BCIC IV 2A: $BAC = 0.0540 \cdot ln(N) -0.1635, (R^{2} = 0.986)$; BCIC2020-3: $BAC = 0.0524 \cdot ln(N) -0.2139, (R^{2} = 0.982)$}
\label{fig-scaling-data}
\end{figure*}
\subsubsection{5.1 Scaling laws with model size}
To analyze how downstream performance scales with model size, we conducted experiments with two additional pre-training settings: a 0.8M-parameter model (4 layers, 128 hidden size, 4 heads) and a 51M-parameter model (12 layers, 512 hidden size, 8 heads), alongside CoMET-Tiny (5M), Base (19M), and Large (151M). We evaluated these five models on the datasets BCIC IV 2A and BCIC2020-3, with the results illustrated in Figure \ref{fig-scaling-model}. On BCIC IV 2A, the balanced accuracy (BAC) scales with model size (M) as: $BAC = 0.013 \cdot ln(M) + 0.568, (R^{2} = 0.949)$. Similarly, on BCIC2020-3, the scaling law is: $BAC = 0.009 \cdot ln(M) + 0.516, (R^{2} = 0.989)$.
\subsubsection{5.2 Scaling laws with data size}
We conducted pre-training experiment on CoMET-Base using 20\%, 40\%, 60\%, 80\% and 100\% of the training data (total 1.91M samples) and tested them on datasets BCIC IV 2A and BCIC2020-3. The results are illustrated in Figure \ref{fig-scaling-data}.  On BCIC IV 2A, the balanced accuracy (BAC) scales with model size (N) as: BAC = 0.0128 · ln(N) + 0.567 ($R^{2}$ = 0.915). Similarly, on BCIC2020-3, the scaling law is: BAC = 0.0231 · ln(N) + 0.512 ($R^{2}$ = 0.933).

\subsection{6. Channel embedding similarity}

\begin{figure*}[!htbp]
\centering
\includegraphics[width=0.8\textwidth]{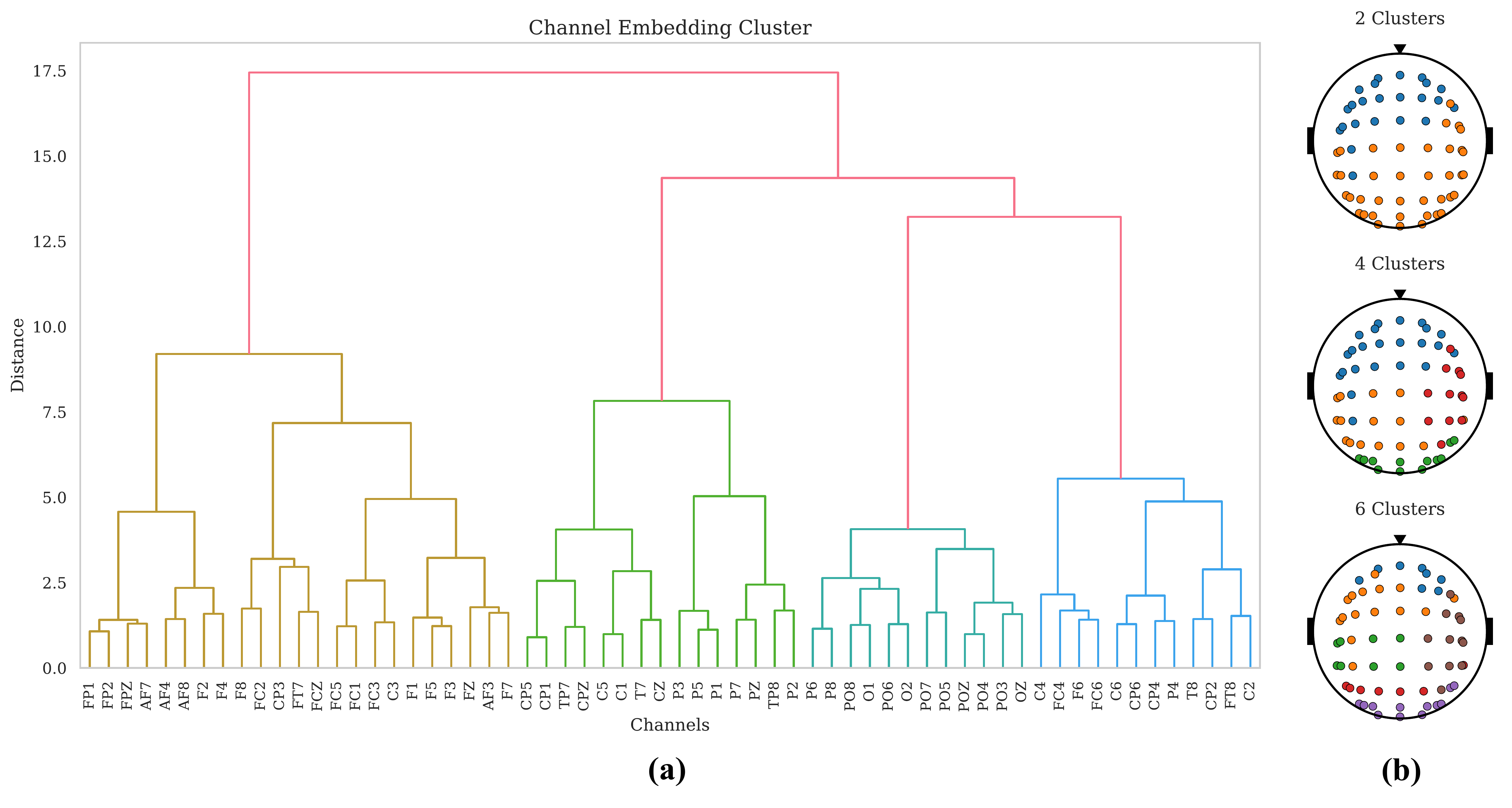}
\caption{The learned EEG channel embeddings cluster of the model CoMET-Tiny. (a): Dendrogram shhows the cosine similarity of channel embeddings. (b): Visualization in topology of the channel embeddings based on the dendrogram results, with the same color representing the same cluster.}
\label{cluster-tiny}
\end{figure*}

\begin{figure*}[!htbp]
\centering
\includegraphics[width=0.8\textwidth]{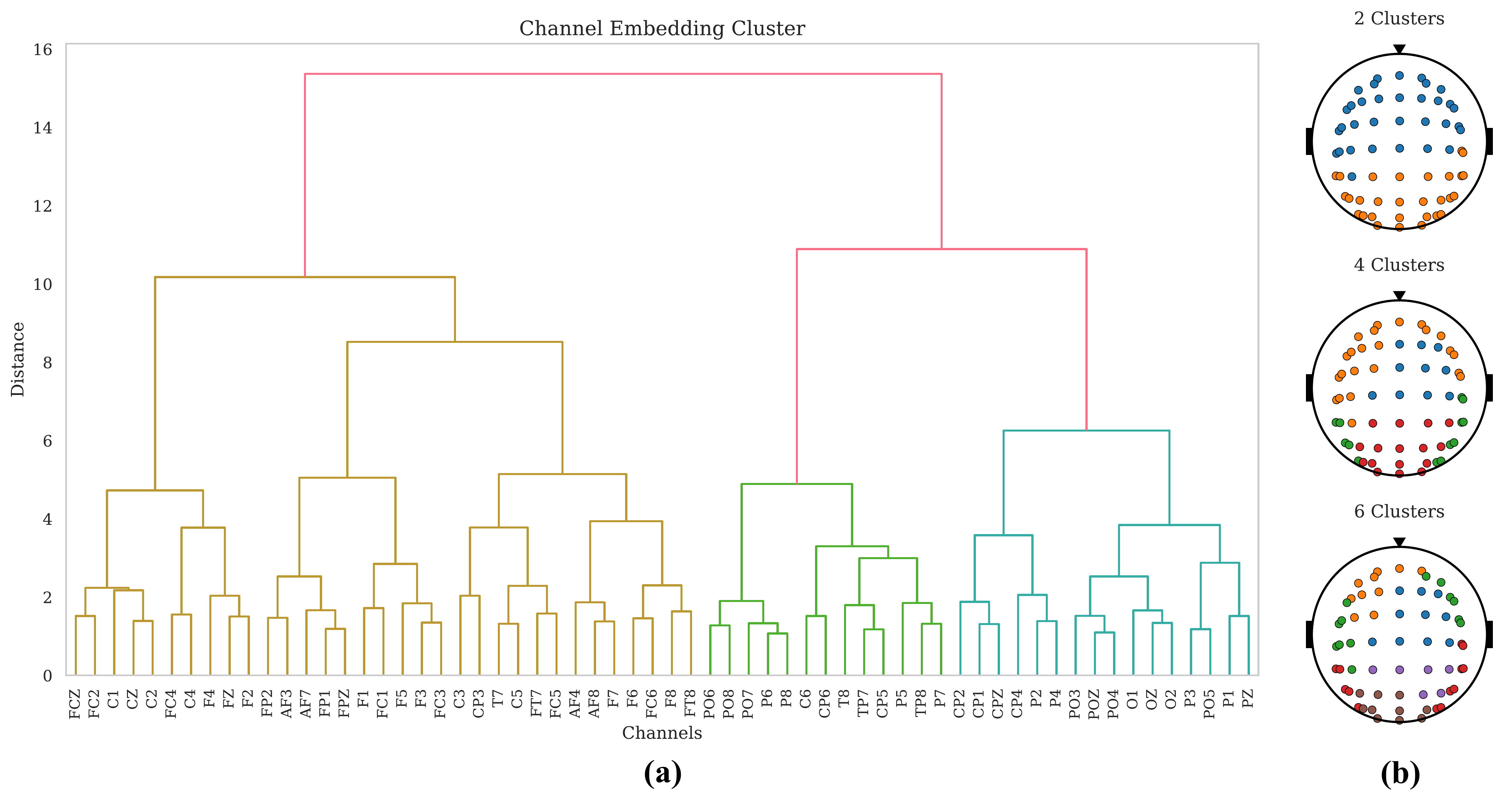}
\caption{The learned EEG channel embeddings cluster of the model CoMET-Base. (a): Dendrogram shhows the cosine similarity of channel embeddings. (b): Visualization in topology of the channel embeddings based on the dendrogram results, with the same color representing the same cluster.}
\label{cluster-base}
\end{figure*}

\begin{figure*}[htbp]
\centering
\includegraphics[width=0.8\textwidth]{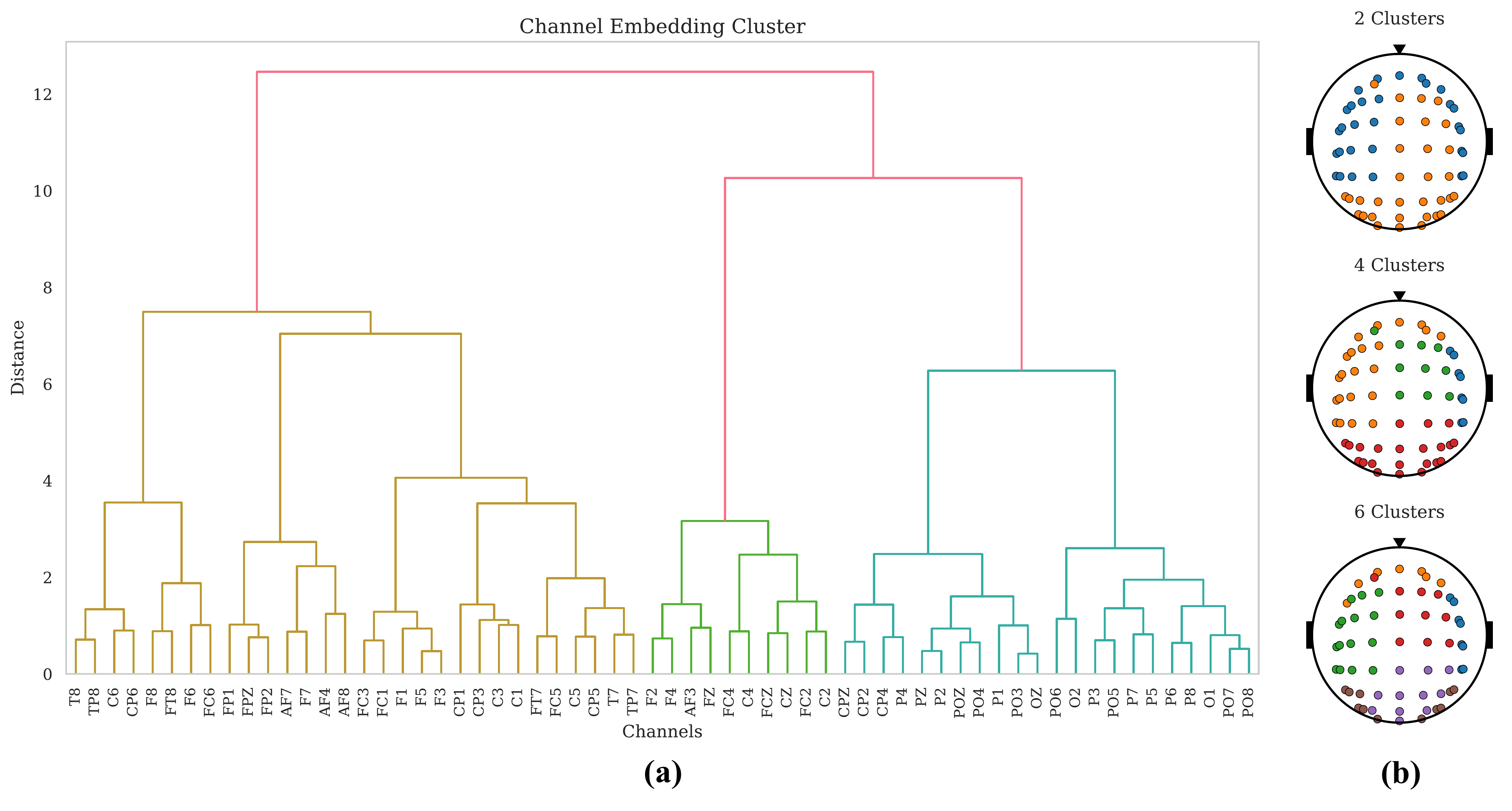}
\caption{The learned EEG channel embeddings cluster of the model CoMET-Large. (a): Dendrogram shhows the cosine similarity of channel embeddings. (b): Visualization in topology of the channel embeddings based on the dendrogram results, with the same color representing the same cluster.}
\label{cluster-large}
\end{figure*}
After pre-training, the foundation models leverage channel embeddings to impose biases on the signals recorded on different channels. Prior work of other large brain models (\cite{jianglarge, wang2024eegpt, wang2025cbramod}) has not, however, visualised the properties of these learned embeddings. We present an exploratory analysis of channel-embedding similarity as design instructions for future brain foundation models. Concretely, we visualised the spatial embeddings by (i) performing hierarchical clustering with cosine similarity as the distance metric and (ii) displaying the resulting channel clusters on EEG topographic maps. The visualisations for CoMET-Tiny, CoMET-Base, and CoMET-Large are reported in Figures \ref{cluster-tiny}, \ref{cluster-base}, and \ref{cluster-large}, respectively.

Across all three models, the embeddings consistently partition into two coarse clusters corresponding to frontal and posterior electrodes. While finer-grained clustering patterns differ slightly between model sizes, they broadly respect canonical neuroanatomical regions—namely the frontal, central, occipital, and temporal lobes. These findings indicate that pre-training enables the models to internalise meaningful EEG spatial structure, and they provide embedding-level evidence that neighbouring electrodes tend to share similar feature representations.

\subsection{7. Feature Distributions}
\begin{figure*}[!htbp]
\centering
\includegraphics[width=0.9\textwidth]{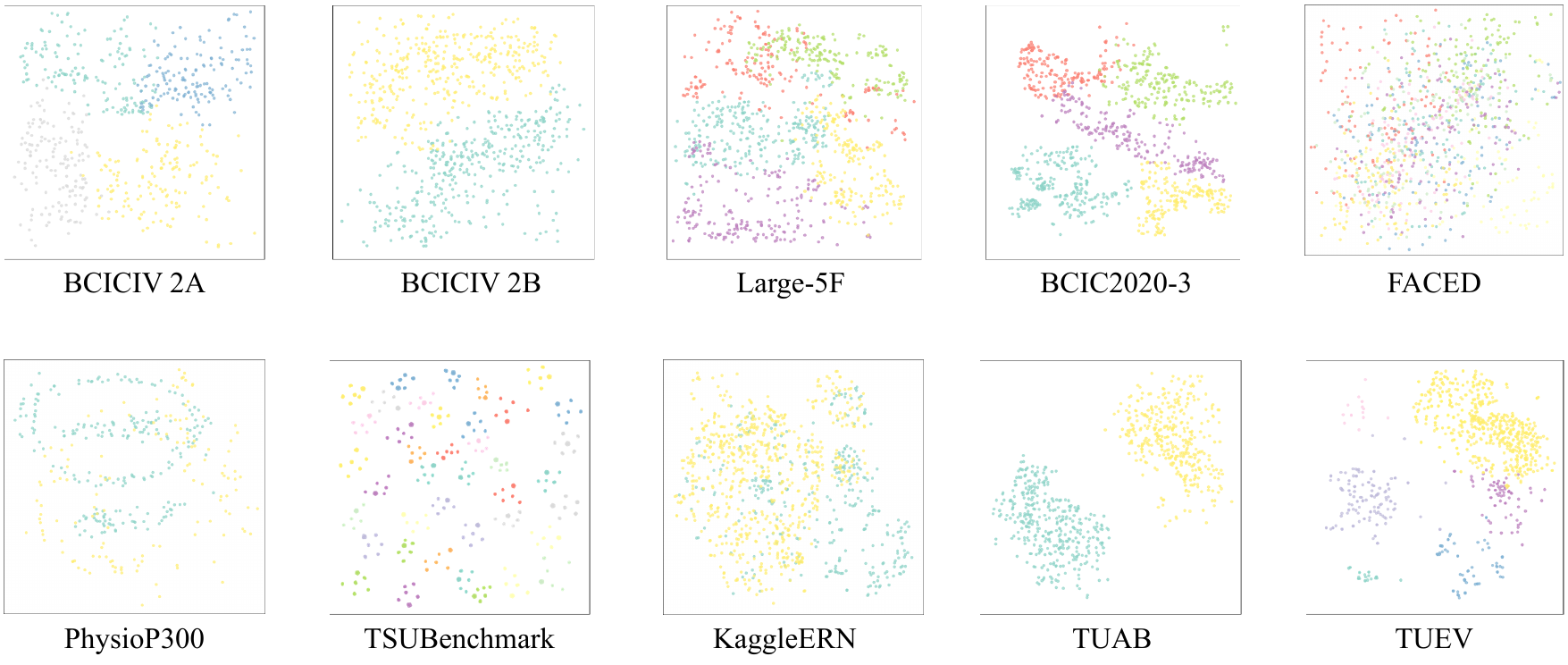}
\caption{T-SNE visualizations of feature distributions on different datasets with CoMET-Large.}
\label{fig:tsne}
\end{figure*}
To showcase the distribution patterns of feature representations, we performed a t-SNE analysis \cite{tSNE}, a widely used nonlinear dimensionality reduction technique, on the learned features across different downstream tasks. This visualization offers an intuitive perspective on how features are organized in the latent space and reveals the clustering behaviour and separability among different classes, as illustrated in Figure~\ref{fig:tsne}.

For datasets with high classification accuracy, such as BCIC-IV-2A (4 classes, 62.75\%), BCIC-IV-2B (2 classes, 73.22\%), THUBenchmark (40 classes, 92.74\%), and TUAB (2 classes, 82.02\%), we observe that inter-class feature distributions exhibit a high degree of separability. In contrast, for datasets with relatively lower classification accuracy, including KaggleERN (2 classes, 58.78\%), Large-5F (5 classes, 38.97\%), and FACED (9 classes, 39.02\%), the feature representations show significant overlap among different classes. This observation is consistent with the corresponding classification accuracies, indicating that the degree of feature separability across classes aligns well with the overall recognition performance.

\subsection{8. Brain map of attention}

\begin{figure*}[!htbp]
    \centering
    \includegraphics[width=0.6\textwidth]{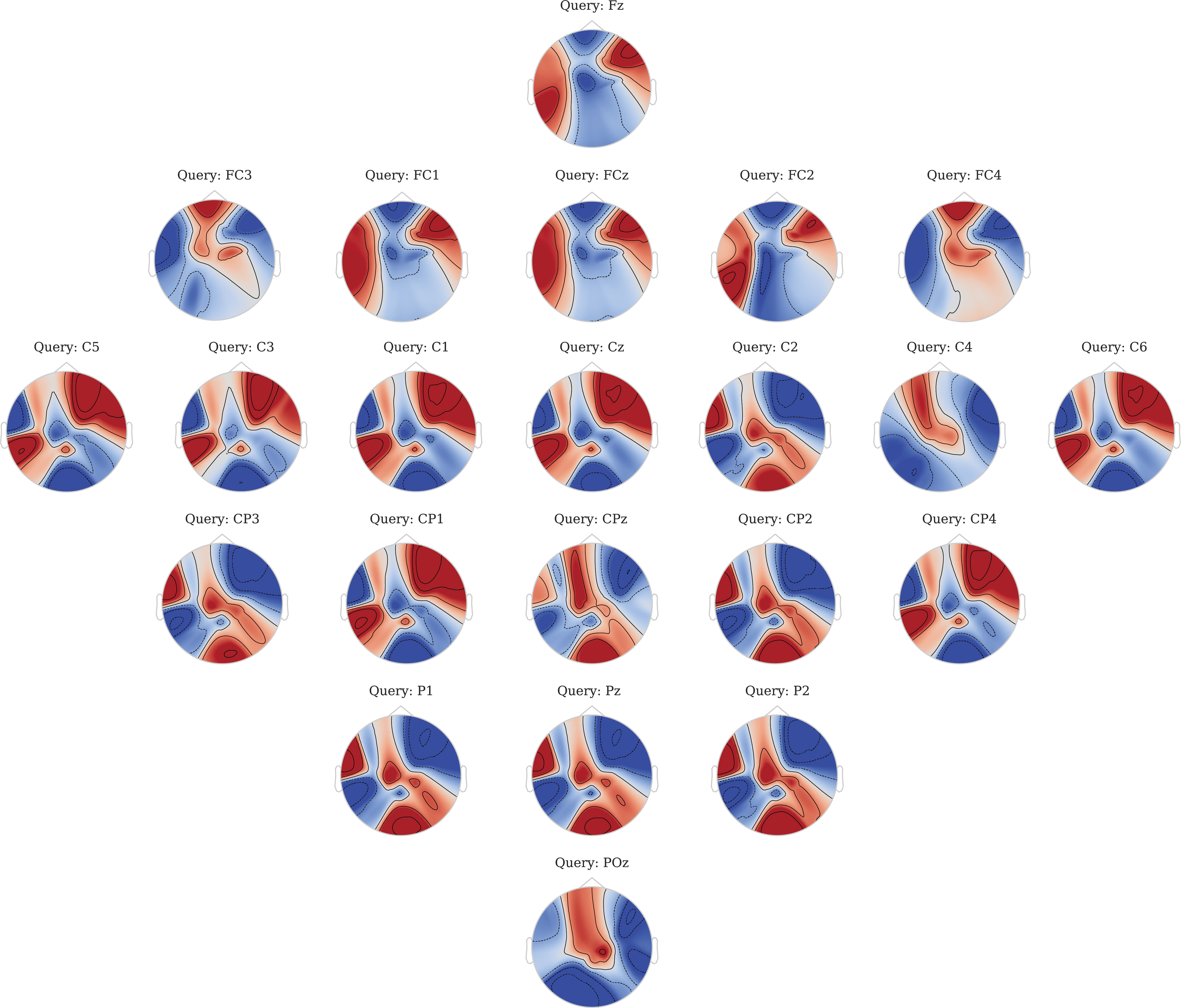}
    \caption{BIOT's attention map for 22 EEG channels (BCIC IV 2A); warmer colors mark stronger attention levels.}
    \label{fig:biot_attn}
\end{figure*}

\begin{figure*}[!htbp]
    \centering
    \includegraphics[width=0.6\textwidth]{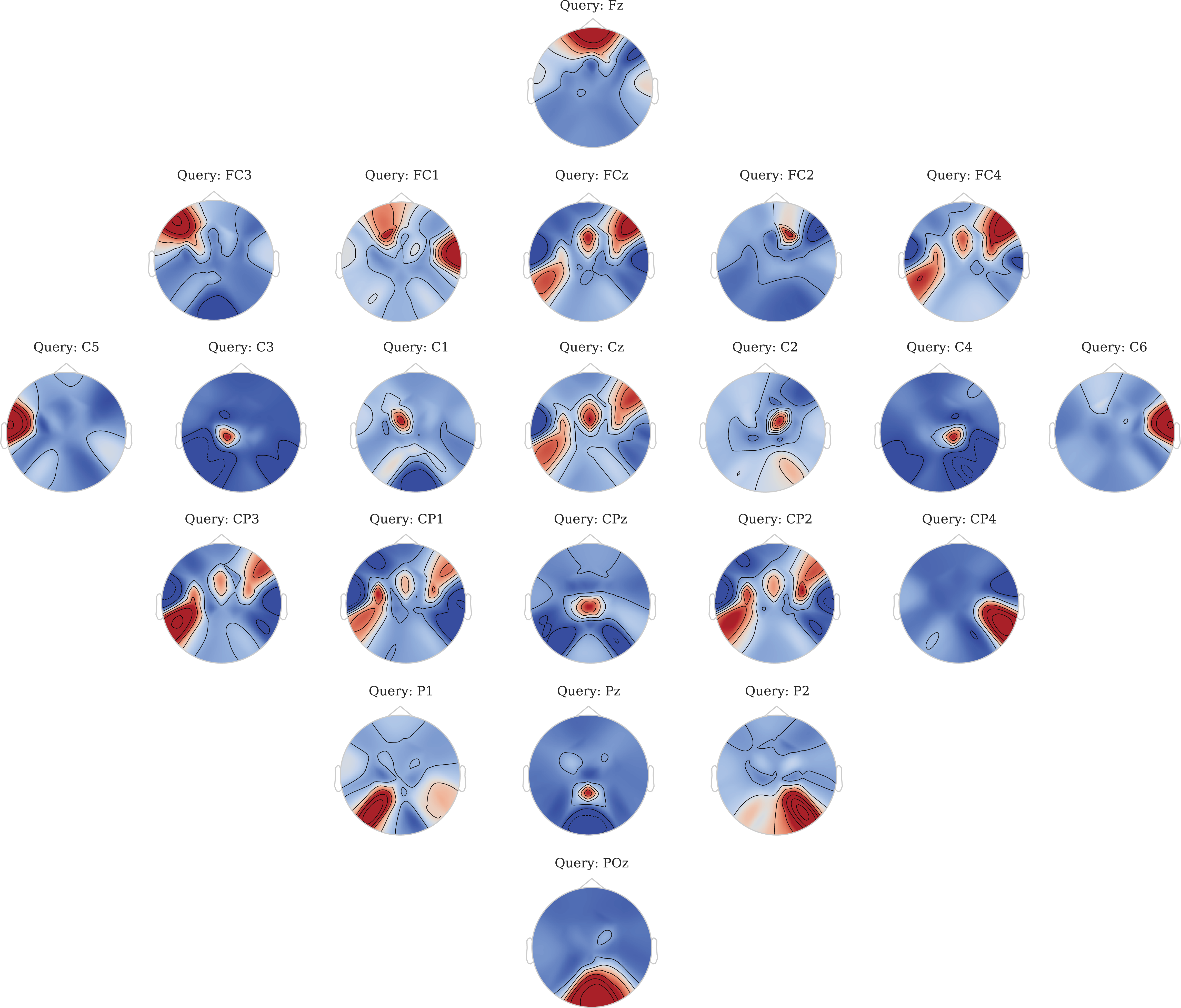}
    \caption{LaBraM's attention map for 22 EEG channels (BCIC IV 2A); warmer colors mark stronger attention levels.}
    \label{fig:labram_attn}
\end{figure*}

\begin{figure*}[!htbp]
    \centering
    \includegraphics[width=0.6\textwidth]{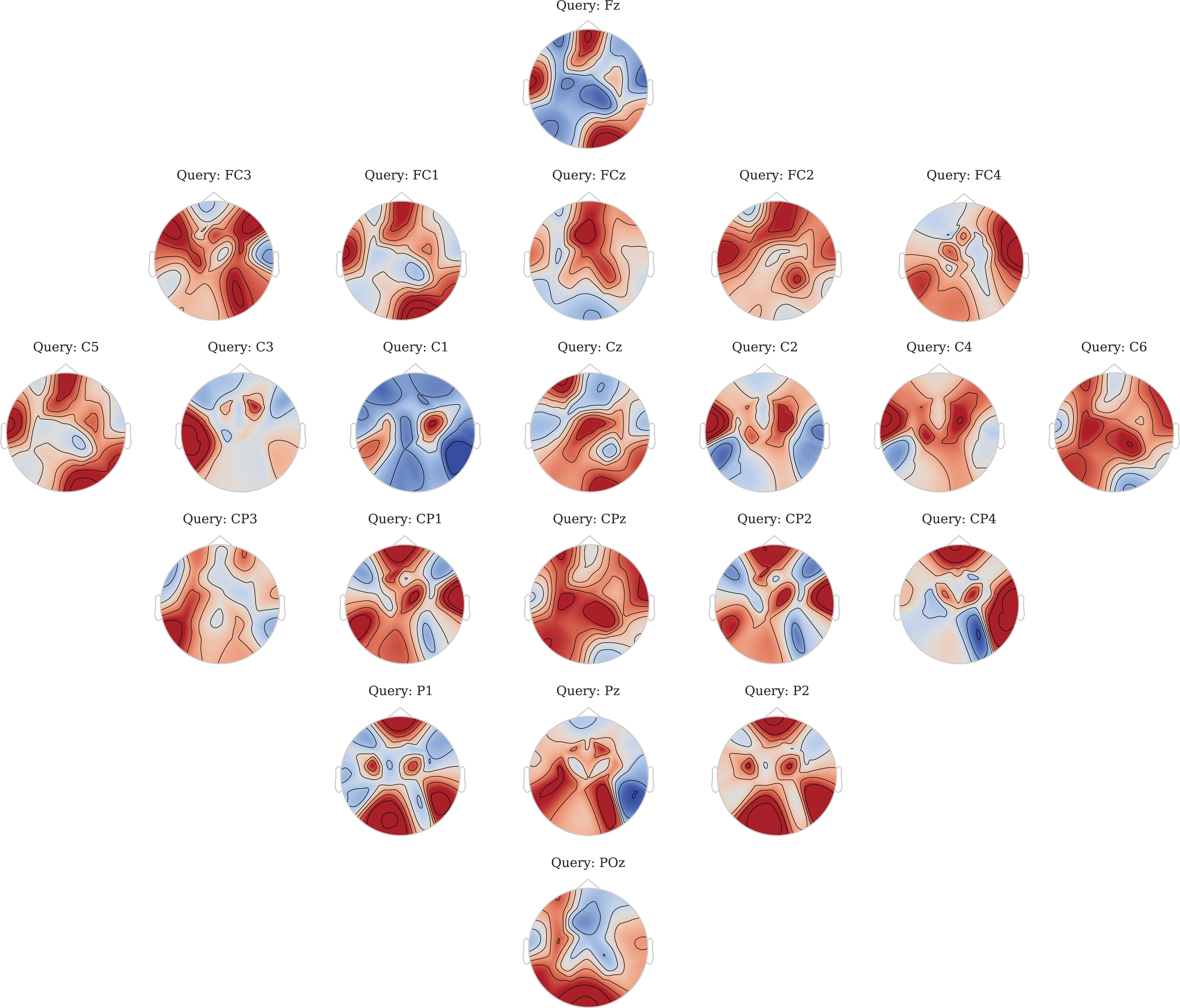}
    \caption{CoMET's attention map for 22 EEG channels (BCIC IV 2A); warmer colors mark stronger attention levels.}
    \label{fig:comet_attn}
\end{figure*}

To facilitate a comprehensive discussion of the model's performance in the EEG paradigm, we present the attention weight distributions across different channels in the BCIC IV 2A dataset (22-channel input) under varying query conditions, as shown in Figure~\ref{fig:biot_attn}, Figure~\ref{fig:labram_attn}, and Figure~\ref{fig:comet_attn}. As observed, when the input query corresponds to different channels, BIOT's attention distribution is restricted to a limited set of paradigms, which exhibit distinct brain region activation patterns in relation to different cognitive tasks. Similarly, LaBram’s attention remains confined to either the current or neighboring channels for most queries, thereby lacking extensive coverage of the brain's diverse regions. In contrast, CoMET demonstrates a much broader attention distribution across channels, regardless of the query input. This characteristic enables CoMET to maintain a high degree of flexibility across varying downstream task paradigms without the need for retraining the encoder, thus positioning it as a highly efficient and generalizable foundational model.

\subsection{9. Different downstream strategies}
In Table~\ref{tab:2a}, Table~\ref{tab:2b}, Table~\ref{tab:ern}, Table~\ref{tab:faced}, Table~\ref{tab:bench}, and Table~\ref{tab:p300}, we present different models alongside CoMET performance under different strategies, namely linear-probing (lb) and fine-tuning (ft), where the former freezes the encoder while the latter fully trains the encoder in the downstream tasks. Some models show significant declines when using strategies different from those in their original paper, such as EEGPT when using the ft strategy and CbraMod when using the lb strategy. In contrast, when using different strategies, the balanced accuracy difference of CoMET is less than 6\%. The lb strategy is generally considered to represent the actual performance of the pre-trained model since the model is not affected by downstream tasks, and CoMET's lb performance is superior than that of other models.

\begin{table*}[!htbp]
\centering
\footnotesize
\begin{tabular}{lcccc}
\toprule
\multirow{2}{*}{\textbf{Model}} 
& \multicolumn{4}{c}{\textbf{BCIC IV 2A}} \\
\cmidrule(lr){2-5}
& \textbf{Strategy} & \textbf{B. Acc} & \textbf{Kappa} & \textbf{F1}\\
\midrule
BIOT  & ft & 41.44 $\pm$ 0.58 & 21.90 $\pm$ 1.19 & 37.37 $\pm$  0.56 \\
  & lb & 42.02 $\pm$ 1.17 & 22.63 $\pm$ 1.56 & 38.10 $\pm$ 1.28 \\
LaBraM & ft & 52.49 $\pm$ 1.34 & 36.60 $\pm$ 2.12 & 52.32 $\pm$ 0.97 \\
    & lb & 51.37 $\pm$ 1.12 & 35.17 $\pm$ 1.48 & 49.73 $\pm$ 1.30 \\
EEGPT & ft & 38.35 $\pm$ 0.31 & 17.69 $\pm$ 0.41 & 37.97 $\pm$ 0.31 \\
     & lb & 51.37 $\pm$ 0.96 & 35.17 $\pm$ 1.26 & 49.73 $\pm$ 0.41 \\
CBraMod   & ft & 55.85 $\pm$ 0.97 & 41.13 $\pm$ 1.30 & 55.08 $\pm$ 1.02 \\
   & lb & 29.04 $\pm$ 0.32 & 5.40 $\pm$ 0.42 & 16.69 $\pm$ 0.46 \\
CoMET  & ft & 57.68 $\pm$ 0.70 & 38.25 $\pm$ 0.93 & 52.81 $\pm$ 0.72\\
  & lb & \textbf{62.75 $\pm$ 1.62} & \textbf{51.70 $\pm$ 1.84} & \textbf{63.37 $\pm$ 1.34}\\
\toprule
\end{tabular}
\caption{Model performance comparisons across different strategies on the BCIC IV 2A Dataset}
\label{tab:2a}
\end{table*}

\begin{table*}[!htbp]
\centering
\footnotesize
\begin{tabular}{lcccc}
\toprule
\multirow{2}{*}{\textbf{Model}}
& \multicolumn{4}{c}{\textbf{BCIC IV 2B}} \\
\cmidrule(lr){2-5}
& \textbf{Strategy} & \textbf{B. Acc} & \textbf{Kappa} & \textbf{F1}\\
\midrule
BIOT  & ft & 50.04 $\pm$ 0.12 & 0.09 $\pm$ 0.25 & 6.31 $\pm$ 0.88 \\
  & lb & 49.79 $\pm$ 4.13 & -0.33 $\pm$ 8.22 & 44.45 $\pm$ 19.62 \\
LaBraM & ft & 54.26 $\pm$ 0.22 & 8.38 $\pm$ 0.43 & 56.86 $\pm$ 1.38 \\
    & lb & 63.93 $\pm$ 7.83 & 27.64 $\pm$ 15.81 & 61.19 $\pm$ 9.96 \\
EEGPT & ft & 59.92 $\pm$ 7.63 & 19.75 $\pm$ 15.40 & 57.45 $\pm$ 8.86 \\
     & lb & 61.07 $\pm$ 0.27 & 22.15 $\pm$ 0.52 & 54.24 $\pm$ 0.48 \\
CBraMod   & ft & 59.46 $\pm$ 0.01 & 18.85 $\pm$ 0.03 & 55.68 $\pm$ 0.03 \\
   & lb & 50.82 $\pm$ 1.86 & 1.68 $\pm$ 3.71 & 40.83 $\pm$ 8.77 \\
CoMET  & ft & \textbf{63.95 $\pm$ 0.84} & \textbf{27.76 $\pm$ 1.68} & \textbf{59.33 $\pm$ 1.47}\\
  & lb & 63.86 $\pm$ 1.78 & 25.05 $\pm$ 2.01 & 61.72 $\pm$ 2.28\\
\toprule
\end{tabular}
\caption{Model performance comparisons across different strategies on the BCIC IV 2B Dataset}
\label{tab:2b}
\end{table*}

\begin{table*}[!htbp]
\centering
\footnotesize
\begin{tabular}{lcccc}
\toprule
\multirow{2}{*}{\textbf{Model}} 
& \multicolumn{4}{c}{\textbf{KaggleERN}} \\
\cmidrule(lr){2-5}
& \textbf{Strategy} & \textbf{B. Acc} & \textbf{Kappa} & \textbf{F1}\\
\midrule
BIOT  & ft & 50.64 $\pm$ 0.01 & 1.62 $\pm$ 0.03 & 71.53 $\pm$ 0.01 \\
  & lb & 52.80 $\pm$ 2.06 & 6.13 $\pm$ 4.19 & 78.11 $\pm$ 4.76 \\
LaBraM & ft & 51.95 $\pm$ 0.81 & 4.38 $\pm$ 1.75 & 77.21 $\pm$ 1.82 \\
    & lb & 57.54 $\pm$ 2.55 & 17.21 $\pm$ 4.47 & 81.10 $\pm$ 1.91 \\
EEGPT & ft & 53.81 $\pm$ 1.87 & 8.85 $\pm$ 3.81 & 79.90 $\pm$ 1.69 \\
     & lb & 54.92 $\pm$ 0.04 & 11.89 $\pm$ 0.12 & 76.77 $\pm$ 0.04 \\
CBraMod   & ft & 53.92 $\pm$ 0.02 & 8.15 $\pm$ 0.05 & 76.49 $\pm$ 0.04 \\
   & lb & 50.57 $\pm$ 0.55 & 1.51 $\pm$ 1.40 & 60.27 $\pm$ 0.98 \\
CoMET  & ft & 58.18 $\pm$ 1.45 & 17.23 $\pm$ 2.63 & 77.93 $\pm$ 2.22\\
  & lb & \textbf{58.78 $\pm$ 0.88} & \textbf{15.24 $\pm$ 0.92} & \textbf{79.66 $\pm$ 1.43}\\
\toprule
\end{tabular}
\caption{Model performance comparisons across different strategies on the Kaggle ERN Dataset}
\label{tab:ern}
\end{table*}

\begin{table*}[!htbp]
\centering
\footnotesize
\begin{tabular}{lcccc}
\toprule
\multirow{2}{*}{\textbf{Model}}
& \multicolumn{4}{c}{\textbf{FACED}} \\
\cmidrule(lr){2-5}
& \textbf{Strategy} & \textbf{B. Acc} & \textbf{Kappa} & \textbf{F1}\\
\midrule
BIOT  & ft & 50.64 $\pm$ 0.01 & 1.62 $\pm$ 0.03 & 71.53 $\pm$ 0.01 \\
  & lb & 13.56 $\pm$ 0.40 & 2.87 $\pm$ 0.46 & 12.28 $\pm$ 0.82 \\
LaBraM & ft & 51.95 $\pm$ 0.81 & 4.38 $\pm$ 1.75 & 77.21 $\pm$ 1.82 \\
    & lb & 21.80 $\pm$ 1.05 & 12.09 $\pm$ 1.23 & 21.57 $\pm$ 1.14 \\
EEGPT & ft & 15.16 $\pm$ 0.06 & 4.67 $\pm$ 0.07 & 15.35 $\pm$ 0.06 \\
     & lb & 54.92 $\pm$ 0.04 & 11.89 $\pm$ 0.12 & 76.77 $\pm$ 0.04 \\
CBraMod   & ft & 53.92 $\pm$ 0.02 & 8.15 $\pm$ 0.05 & 76.49 $\pm$ 0.04 \\
   & lb & 29.06 $\pm$ 1.18 & 19.88 $\pm$ 1.25 & 27.76 $\pm$ 0.73 \\
CoMET  & ft & 40.10 $\pm$ 2.25 & 32.47 $\pm$ 2.52 & 40.13 $\pm$ 2.15\\
  & lb & \textbf{58.78 $\pm$ 0.88} & \textbf{15.24 $\pm$ 0.92} & \textbf{79.66 $\pm$ 1.43}\\
\toprule
\end{tabular}
\caption{Model performance comparisons across different strategies on the FACED Dataset}
\label{tab:faced}
\end{table*}

\begin{table*}[!htbp]
\centering
\footnotesize
\begin{tabular}{lcccc}
\toprule
\multirow{2}{*}{\textbf{Model}}
& \multicolumn{4}{c}{\textbf{THUBenchmark}} \\
\cmidrule(lr){2-5}
& \textbf{Strategy} & \textbf{B. Acc} & \textbf{Kappa} & \textbf{F1}\\
\midrule
BIOT  & ft & 74.17 $\pm$ 0.06 & 73.50 $\pm$ 0.07 & 74.24 $\pm$ 0.05 \\
  & lb & 80.07 $\pm$ 0.89 & 78.54 $\pm$ 0.12 & 80.20 $\pm$ 0.39 \\
LaBraM & ft & 91.43 $\pm$ 0.13 & 91.21 $\pm$ 0.13 & 91.43 $\pm$ 0.13 \\
    & lb & 92.02 $\pm$ 0.92 & 94.79 $\pm$ 0.45 & 95.02 $\pm$ 0.55 \\
EEGPT & ft & 30.23 $\pm$ 0.03 & 28.44 $\pm$ 0.03 & 30.87 $\pm$ 0.03 \\
     & lb & 82.59 $\pm$ 0.09 & 82.15 $\pm$ 0.09 & 82.57 $\pm$ 0.08 \\
CBraMod   & ft & 91.45 $\pm$ 0.24 & 91.23 $\pm$ 0.25 & 91.43 $\pm$ 0.24 \\
   & lb & 33.04 $\pm$ 0.92 & 27.88 $\pm$ 0.90 & 32.19 $\pm$ 0.18 \\
CoMET  & ft & \textbf{96.88 $\pm$ 0.94} & \textbf{96.63 $\pm$ 0.11} & \textbf{96.86 $\pm$ 0.14}\\
  & lb & 92.74 $\pm$ 1.61 & 92.65 $\pm$ 2.46 & 93.50 $\pm$ 0.81\\
\toprule
\end{tabular}
\caption{Model performance comparisons across different strategies on the THUBenchmark dataset}
\label{tab:bench}
\end{table*}

\begin{table*}[!htbp]
\centering
\footnotesize
\begin{tabular}{lcccc}
\toprule
\multirow{2}{*}{\textbf{Model}}
& \multicolumn{4}{c}{\textbf{PhysioP300}} \\
\cmidrule(lr){2-5}
& \textbf{Strategy} & \textbf{B. Acc} & \textbf{Kappa} & \textbf{F1}\\
\midrule
BIOT  & ft & 50.04 $\pm$ 0.12 & 0.09 $\pm$ 0.25 & 6.31 $\pm$ 0.88 \\
  & lb & 49.79 $\pm$ 4.13 & -0.33 $\pm$ 8.22 & 44.45 $\pm$ 19.62 \\
LaBraM & ft & 54.26 $\pm$ 0.22 & 8.38 $\pm$ 0.43 & 56.86 $\pm$ 1.38 \\
    & lb & 61.93 $\pm$ 7.83 & 27.64 $\pm$ 15.81 & 61.19 $\pm$ 9.96 \\
EEGPT & ft & 59.92 $\pm$ 7.63 & 19.75 $\pm$ 15.40 & 57.45 $\pm$ 8.86 \\
     & lb & 61.07 $\pm$ 0.27 & 22.15 $\pm$ 0.52 & 54.24 $\pm$ 0.48 \\
CBraMod   & ft & 59.46 $\pm$ 0.01 & 18.85 $\pm$ 0.03 & 55.68 $\pm$ 0.03 \\
   & lb & 50.82 $\pm$ 1.86 & 1.68 $\pm$ 3.71 & 40.83 $\pm$ 8.77 \\
CoMET  & ft & \textbf{63.95 $\pm$ 0.84} & \textbf{27.76 $\pm$ 1.68} & \textbf{59.33 $\pm$ 1.47}\\
  & lb & 63.86 $\pm$ 1.78 & 25.05 $\pm$ 2.01 & 61.72 $\pm$ 2.28\\
\toprule
\end{tabular}
\caption{Model performance comparisons across different strategies on the PhysioP300 Dataset}
\label{tab:p300}
\end{table*}

\end{document}